\documentclass[prd,aps,superscriptaddress,nofootinbib,amsmath,amssymb,showpacs,9pt]{revtex4}
\usepackage{graphicx}
\usepackage{braket}
\usepackage{nicefrac}
\usepackage{float}%
\usepackage{amsmath}
\usepackage{hyperref}
\usepackage{multirow}
 \usepackage{bbold}
 \usepackage{slashed}
 \usepackage{graphics}
\usepackage{graphicx}
\usepackage{dcolumn}
\usepackage{bm}
\usepackage{multirow}
\usepackage{adjustbox}
\usepackage{textcomp}

\usepackage[usenames, dvipsnames]{color}

\def\slashchar#1{\setbox0=\hbox{$#1$}
   \dimen0=\wd0 \setbox1=\hbox{/} \dimen1=\wd1
   \ifdim\dimen0>\dimen1 \rlap{\hbox to \dimen0{\hfil/\hfil}} #1
   \else  \rlap{\hbox to \dimen1{\hfil$#1$\hfil}} / \fi}

\graphicspath{{./eps/}}

\begin{document}
\title{Charged current induced electron-proton scattering and the axial vector form factor}

\author{A. \surname{Fatima}}
\affiliation{Department of Physics, Aligarh Muslim University, Aligarh-202 002, India}
\author{M. Sajjad \surname{Athar}}
\email{sajathar@gmail.com}
\affiliation{Department of Physics, Aligarh Muslim University, Aligarh-202 002, India}
\author{S. K. \surname{Singh}}
\affiliation{Department of Physics, Aligarh Muslim University, Aligarh-202 002, India}

\begin{abstract} 

We investigate the total scattering cross section~($\sigma$), the differential cross section~$\left(\frac{d\sigma}{dQ^2}\right)$, the longitudinal~($A_L(E_e,Q^2)$) and  perpendicular~($A_P(E_e,Q^2)$) spin asymmetries of the polarized target proton, as well as the longitudinal~($P_L(E_e,Q^2)$),  perpendicular~($P_P(E_e,Q^2)$), and transverse~($P_T(E_e,Q^2)$) polarization components of the final neutron, in the weak charged current induced electron-proton scattering relevant to the { future} experiments at the Thomas Jefferson National Accelerator Facility~(JLab) and Mainz Microtron~(MAMI). The analysis is performed assuming time-reversal~(T) invariance as well as without assuming T invariance, allowing for a nonvanishing transverse polarization component of the final nucleon, perpendicular to the production plane. Numerical results are presented for { the above mentioned} observables, and their sensitivities to the { various parameterizations of the axial vector form factor $g_1(Q^2)$ and a nonzero weak electric form factor $g_2(Q^2)$} are examined. { We find that the cross section depends strongly on the parameterizations used for the axial vector form factor. Moreover, the dipole parameterization of $g_1(Q^2)$ with a higher value of the axial dipole mass $M_A$ simulates the apparent enhancement in $\sigma$ obtained using the non-dipole parameterizations like the $z$-expansion and Faddeev equation form. 
The cross sections are found to depend only weakly on the weak electric form factor $g_2(Q^2)$, which is associated with the violation of G-invariance. On the contrary, the spin observables both $A_{L,P} (E_e, Q^2)$ and $P_{L,P} (E_e, Q^2)$ are found to be strongly dependent on $g_2(Q^2)$.}
This study may be useful in the { analysis of the} neutrino oscillation experiments to provide an alternative constrain on the parameterization of axial vector form factor, which currently has large uncertainties.
\end{abstract}
\maketitle

\section{Introduction}
Electron scattering on the polarized proton targets constitutes a powerful and uniquely sensitive probe of the weak interactions, nucleon structure, and fundamental symmetries. High-precision electron-proton scattering from nucleon and nuclear target experiments conducted at the Thomas Jefferson National Accelerator Facility~(JLab)~\cite{Adderley:2022uql} and MIT-Bates~\cite{Beise:2004py, BatesFPP:1997rpw} in the United States, and Mainz Microtron~(MAMI)~\cite{Maas:2006} in Germany have played a pivotal role in providing stringent low-energy tests of the Standard Model, while continuously advancing the frontiers of electroweak physics. Unlike strong and electromagnetic interactions, weak interactions violate parity, thereby enabling access to observables that are otherwise inaccessible in scattering experiments where parity is conserved. In particular, the parity-violating asymmetries measured in polarized electron–proton scattering allow the isolation of weak neutral current effects with exceptional sensitivity. 
In the region of low electron energies, the experiments at MAMI and MIT-Bates and the experiments at JLab in the higher energy region of electrons employing highly polarized electron beams have made significant contributions to precision measurements in low-energy electroweak physics through dedicated parity-violation studies. In recent times, efforts at JLab and MAMI have further strengthened these investigations by achieving unprecedented control over systematic uncertainties. Looking ahead, forthcoming experiments such as the upgraded facilities at JLab and the P2 experiment at MAMI promise substantial improvements in precision, offering enhanced sensitivity to possible deviations from Standard Model predictions and opening new windows for the discovery of physics beyond the Standard Model~\cite{Kumar:2013yoa}.

In describing the weak interactions of electrons and neutrinos with nucleon targets, nucleon form factors are fundamental quantities that are used to describe the internal structure of the nucleon and play a central role in our understanding of nonperturbative QCD dynamics. The vector component of the weak processes is constrained using the isovector combination of the proton and neutron electromagnetic form factors, based on the isospin symmetry and the conserved vector current~(CVC) hypothesis~\cite{SajjadAthar:2022pjt}, and are obtained through the experimental measurements of electric~($G_E(Q^2)$) and magnetic~($G_M(Q^2)$) Sachs' form factors. In contrast, the axial vector contribution, parameterized by the axial vector~($g_1(Q^2)$) and induced pseudoscalar~($g_3(Q^2)$) form factors, is comparatively less well determined. The isovector axial vector form factor of the nucleon occupies a particularly important position as the pseudoscalar form factor $g_{3}(Q^2)$ is generally expressed in terms of $g_1 (Q^2)$ using the hypothesis of partially conserved axial vector current~(PCAC) hypothesis and pion pole dominance of the divergence of the axial vector current. 
The value of the axial vector form factor at zero momentum transfer defines the nucleon axial charge,  which is primarily determined from neutron beta decay to be $g_1(0)=1.267 \pm 0.0030$~\cite{ParticleDataGroup:2024cfk}, and is one of the most precisely measured quantities in hadronic physics~\cite{ParticleDataGroup:2024cfk, Gonzalez-Alonso:2018omy}. Moreover, the momentum-transfer dependence of the axial vector form factor is of crucial importance for a wide range of weak interaction processes at higher energies~(momentum transfers), most notably for the quasielastic neutrino-nucleon scattering. In recent years, several studies have emphasized  that uncertainties in the measurement of the axial vector form factor directly propagate into systematic errors in the extraction of neutrino oscillation parameters, making a precise theoretical understanding of $g_1(Q^2)$  indispensable for the current and future neutrino oscillation experiments~\cite{Ankowski:2022thw}.

Traditionally, the axial vector form factor has been extracted using a combination of neutrino-induced charged-current quasielastic scattering on the nuclear targets including light nuclei like deuterium~\cite{Bodek:2007vi}, threshold pion electroproduction data from proton~\cite{Muller:2003jz}, and weak processes such as muon capture in hydrogen~\cite{MuCap:2012lei, MuCap:2015boo, Czarnecki:2007th}. While these methods have provided valuable insights, they suffer from experimental limitations, nuclear corrections, and model-dependent assumptions, leading to persistent uncertainties and tensions among different datasets~\cite{Alvarez-Ruso:2022ctb}. In general, the axial vector form factor is parameterized to be of the dipole form, where the choice of the numerical value of the axial dipole mass $M_A$ plays a crucial role in determining the neutrino-nucleon quasielastic~(QE) cross section and has been a subject of intense and sustained debate within the neutrino physics community over the past several years. A remarkably wide range of values for $M_A$,  has been reported and discussed in literature, reflecting both experimental and theoretical uncertainties as well as differences in data interpretation~\cite{SajjadAthar:2022pjt, Tomalak:2026wsu}.

The early measurements of $M_A$ were performed using neutrino and antineutrino beams incident on heavy nuclear targets~\cite{Bernard:2001rs}. These experiments were primarily carried out in the 1970s and 1980s and relied on bubble chambers and early tracking detectors using Freon~($CF_3Br$),  Freon–propane mixtures~($CF_3Br-C_3H_8$),  Iron~(Fe), Neon~(Ne), and Aluminum~(Al), which were exposed to wide-band neutrino beams at CERN and other facilities. CERN experiments~\cite{Bonetti:1977cs, GARGAMELLENEUTRINOPROPANE:1979kqk, Armenise:1979zg} provided early values around $M_A=1.0$~GeV but suffered from large systematic errors due to limited statistics. SKAT data~\cite{Belikov:1983kg, Asratian:1984gh, Grabosch:1986js} analyzed primarily high-$Q^2$ regions to reduce nuclear uncertainties. In order to reduce the large uncertainty arising due to the nuclear medium, the experiments at ANL~\cite{Barish:1977qk, Radecky:1981fn}, BNL~\cite{Baker:1981su, Kitagaki:1983px} using hydrogen and deuterium targets, determined the value of $M_{A}$ by using a dipole form of the axial vector form factor. A reanalysis of these classic datasets by Bodek et al.~\cite{Bodek:2007vi}  yielded a relatively precise value of $M_A= 1.014 \pm 0.014$~GeV. In contrast, Meyer et al.~\cite{Meyer:2016oeg}, analyzing the same experimental data from the ANL~\cite{Barish:1977qk, Radecky:1981fn} and BNL~\cite{Baker:1981su, Kitagaki:1983px} experiments, reported a broader range of values, $M_A= 1.02-1.17$~GeV, depending on which dataset was included in the fit. An earlier comprehensive study by Bernard et al.~\cite{Bernard:2001rs}, which combined neutrino and antineutrino scattering data on hydrogen and deuterium with threshold pion electroproduction data from proton, obtained the best fit at $M_A= 1.026 \pm 0.021$~GeV.

With the advent of modern high-statistics experiments, a large body of new QE neutrino-nucleus scattering data have become available, covering both low- and intermediate-energy regimes and involving a variety of nuclear targets. Analyses of data from the NOMAD~\cite{NOMAD:2009qmu} and MINERvA~\cite{MINERvA:2018hqn, MINERvA:2013kdn} experiments, performed at relatively higher neutrino energies, continue to favor a lower value of $M_A$, close to 1.03 GeV, consistent with the earlier free-nucleon results. In sharp contrast, several experiments operating predominantly in the few hundreds of MeV to GeV energy range and using nuclear targets, including MiniBooNE~\cite{MiniBooNE:2007iti}, MINOS~\cite{MINOS:2014axb}, K2K~\cite{K2K:2006odf}, T2K~\cite{T2K:2014hih}, and SciBooNE~\cite{Alcaraz-Aunion:2009pzm}, report significantly larger effective values of $M_A$, typically in the range 1.2--1.35 GeV. This apparent discrepancy between low energy and high energy measurements has emerged as one of the central puzzles in the interpretation of QE neutrino scattering data.
One possible explanation for the larger values of $M_A$ extracted at lower energies is the inadequate treatment of nuclear medium effects, such as multinucleon correlations, meson-exchange currents, and final-state interactions, which can enhance the observed QE-like cross section and mimic the effect of a larger axial mass. Since the NOMAD and MINERvA experiments operate at comparatively higher energies, where such nuclear effects are expected to be less pronounced, their consistency with the lower $M_A$ value lends support to this interpretation. This issue has been actively discussed in recent literature~\cite{SajjadAthar:2022pjt}.

A major recent advance in this area has been achieved by the MINERvA collaboration~\cite{MINERvA:2023avz}, which performed the first high-statistics determination of $g_1(Q^2)$ from free-nucleon data using antineutrino scattering on hydrogen embedded in a plastic scintillator detector. This measurement represents a significant milestone in the study of $g_{1} (Q^2)$. MINERvA measured the differential cross section as a function of the squared four-momentum transfer, $Q^2$, and from these data extracted both the axial vector form factor $g_1(Q^2)$ and the axial radius $r_A$, which characterizes the spatial distribution of axial charge. By minimizing nuclear effects, this work provides a cleaner experimental benchmark for nucleon axial structure. 

{ On the theoretical side, there have been many calculations for the determination of $g_1(Q^2)$ by studying the total~($\sigma(E)$) and the differential~$\left(\frac{d\sigma (E)}{dQ^2}\right)$ cross sections and their dependence on the axial dipole mass $M_A$ in the scattering of (anti)neutrinos from the nucleons and nuclei~\cite{Akbar:2015yda, Benhar:2009wi, Benhar:2010nx, NuSTEC:2017hzk, Kuzmin:2007kr, SajjadAthar:2022pjt}. In a recent study, Alvarez-Ruso et al.~\cite{Alvarez-Ruso:2018rdx} employed a Bayesian neural-network framework to extract the nucleon axial form factor in a model-independent way from neutrino-deuteron ANL data. They revealed a striking sensitivity to low-$Q^2$ data and deuteron corrections, leading to tensions in the extracted axial radius, and underscored the need for high-precision neutrino quasielastic measurements on hydrogen and deuterium.
The possibility of determining $M_A$ by studying the various spin observables of the nucleons in the (anti)neutrino scattering from nucleons induced by the weak charged and neutral currents has  been explored earlier by Block~\cite{Block:1964gj} and later by Bilenky and Christova~\cite{Bilenky:2013fra, Bilenky:2013iua}.
In recent years, Tomalak~\cite{Tomalak:2020zlv} investigated the neutrino scattering on the polarized and unpolarized nucleons, including the recoil polarization observables, and demonstrated that the parity-violating weak interaction induces sizable spin asymmetries at the leading order. Such measurements provide an independent access to the proton axial structure, enabling, in particular, the first direct extraction of the pseudoscalar form factor from neutrino data without relying on the PCAC ansatz or pion-pole dominance assumptions. Graczyk and Kowal~\cite{Graczyk:2019xwg} demonstrated that single-, double-, and triple- spin asymmetries in quasielastic charged current (anti)neutrino scattering are highly sensitive probes of the nucleon axial vector form factor. In particular, the target-recoil and lepton-target-recoil asymmetries exhibit a strong dependence on the axial dipole mass, while all asymmetries, except the lepton polarization, provide observables for probing the second class current contributions. }

In parallel, substantial progress has been made on the theoretical front through lattice QCD, which offers a powerful and systematically improvable framework for computing nucleon axial properties directly from the underlying theory of the strong interaction. Over the past two decades, lattice determinations of the axial charge $g_1(0)$ have improved dramatically. Recent simulations employ near-physical pion masses, large lattice volumes, and advanced analysis techniques such as multi-state fits and variational methods, leading to results that are increasingly consistent with experimental determinations. Extending beyond the axial charge, lattice QCD enables a first-principles determination of the $Q^2$ dependence of $g_1(Q^2)$ over a range of spacelike momentum transfers, providing direct input for neutrino-nucleon cross sections without relying on phenomenological parameterizations such as the dipole ansatz~\cite{Gupta:2024qip, Jang:2023zts, Petti:2023abz, Tomalak:2026wsu}.

Recent studies have analyzed the axial vector form factor using both traditional dipole fits as well as using lattice QCD with the more general z-expansion~\cite{Hill:2010yb}, while the extrapolation of lattice results to the physical limit has been guided by expressions derived within the chiral perturbation theory (ChPT)~\cite{Alvarez-Ruso:2025oak}. Moreover, lattice QCD has contributed to the study of related axial vector parameters, including the induced pseudoscalar form factor and the pion-nucleon coupling through the axial Ward-Takahashi identity, thereby reinforcing the internal theoretical consistency of the framework~\cite{Aoki:2025taf}. Yao et al.~\cite{Yao:2017fym} have considered both the pion mass and momentum transfer dependences of the nucleon axial vector form factor in baryon chiral perturbation theory (BChPT) and investigated their dependences by performing fits to the recent lattice QCD data  without and with explicit $\Delta(1232)$ contribution. The synergy between high-precision experimental measurements by MINERvA and first-principles lattice QCD calculations has led to significant progress in the last one decade. In particular, combined analyses of MINERvA data and lattice QCD results have enabled meaningful and quantitative comparisons between experiment and theory, shedding new light on the momentum dependence of the nucleon axial vector form factor and the value of the axial radius~\cite{Meyer:2026kdl, MINERvA:2025ygc}. These developments mark an important step toward a unified and precise understanding of the nucleon axial structure, with far-reaching implications for the hadron structure studies and the precision physics goals of the modern neutrino experiments.

An alternative and potentially more powerful approach for extracting detailed information on the $Q^2$ dependence of the  axial vector form factor $g_1(Q^2)$ may be provided by the weak charged current interaction induced electron scattering from both unpolarized and polarized proton targets. Unlike neutrino induced processes, which suffer from broad energy spectra, comparatively very low event rates, etc., and thus containing large error bars, electron scattering experiments benefit from highly monochromatic beams with exceptionally high luminosity, several orders of magnitude larger than those attainable in neutrino experiments, thereby providing opportunity for the precise measurements with reduced statistical uncertainties and improved control over kinematic reconstruction. Furthermore, the availability of high-purity hydrogen targets in electron scattering experiments offers a clean experimental environment, in contrast to the neutrino experiments where the use of free hydrogen or deuterium targets is severely constrained by security related limitations. These advantages make weak charged current electron proton scattering a particularly promising alternative for probing the nucleon axial structure. 
Recently, { Klest~\cite{Klest:2025bfl}, and independently Yang and Kumar~\cite{Yang:2026vuf}, have demonstrated that at very high energies corresponding to the center of energy $\sqrt{s}=140$~GeV, relevant to the proposed Electron Ion Collider~(EIC) at BNL, and Electron Ion Collider in China~(EIcC), the electron scattering on a longitudinally polarized proton target induced by the weak charged currents offers an avenue to constrain the nucleon axial-vector form factor. They emphasize that target spin asymmetries in the low-$Q^2$ region, where the axial coupling overwhelmingly dominates, constitute a sensitive probe.
However, in the region of moderately high energy corresponding to a few GeV, } an experimental proposal is already under consideration at JLab~\cite{Averett} using polarized electron beam of energy 1.1 and 2.2~GeV to { determine the axial vector form factor in the measurement of} the differential cross section $\frac{d\sigma}{d\Omega_{\nu}}$ around $Q^2 \approx 1$~GeV$^{2}$, where it has maximum uncertainty. The experiment is feasible provided the background due to the electromagnetic electron-proton scattering and pion photoproduction could be kept under control~\cite{Averett}.  Moreover, a similar experiment using polarized positron beam on the deuterium target to extract the axial vector form factor is also under consideration at JLab~\cite{Androic}.

{ The present work is motivated by the aforementioned proposals~\cite{Averett, Androic} to extend our earlier works~\cite{Fatima:2018gjy, Akbar:2016awk, Akbar:2017qsf, Fatima:2018tzs, Fatima:2018wsy, Fatima:2020pvv, Fatima:2022tlf} on the study of polarization observables of the final nucleon/hyperon and lepton in the antineutrino~\cite{Akbar:2016awk, Fatima:2018tzs, Fatima:2018wsy, Fatima:2020pvv, Fatima:2022tlf} as well as the electron~\cite{Akbar:2017qsf,Fatima:2018gjy} induced scattering from the free nucleon and nuclear targets, to perform a comprehensive study of the weak charged current induced electron-proton scattering, focusing on a wide range of observables relevant to the ongoing and future experiments at JLab, which can also be done, in principle, at MAMI. 
In Refs.~\cite{Akbar:2017qsf, Fatima:2018gjy}, we developed the formalism, following the works of Bilenky and Christova~\cite{Bilekny, Bilenky:2013fra, Bilenky:2013iua}, to study the electron induced hyperon/nucleon production from the free nucleon target and the polarization observables of the final hyperons/nucleon, assuming T invariance as well as its violation. 
In this work, we have used the earlier developed formalism to study the polarization observables of the final nucleon in the weak electron-proton scattering, specifically for the kinematical observables relevant for the JLab proposals~\cite{Averett, Androic}. Furthermore, we have also studied, the spin asymmetries of the polarized target proton. The sensitivity of the cross sections, spin asymmetries of the polarized target, and the polarization observables of the final nucleon to the axial vector form factor has been studied by using the modern alternatives of the traditional dipole parameterization like the Chen-Roberts~\cite{Chen:2022odn} Faddeev equation form and the four $z$-expansion fits of Meyer et al.~\cite{MINERvA:2025ygc}, namely old deuterium data fit, lattice QCD~(LQCD) calculation fit, MINERvA hydrogen data fit and the combined MINERvA hydrogen and LQCD fit. }

Weak electron-proton scattering experiments, which can be done at MAMI and JLab would provide a high precision, nuclear effect free, and theoretically clean avenue to study the axial structure of the proton.  By directly probing the weak axial vector form factor through the electron scattering experiments, one can significantly improve our knowledge of the axial vector form factor $g_1 (Q^2)$, which would play a decisive role in resolving long-standing discrepancies in neutrino cross section measurements. Specifically, we obtain the total scattering cross section~($\sigma$), differential cross section~($d\sigma/dQ^2$), the longitudinal~($A_L(E_e,Q^2)$) and perpendicular~($A_P(E_e,Q^2)$) spin asymmetries of the polarized proton target, as well as the longitudinal~($P_L(E_e,Q^2)$), perpendicular~($P_P(E_e,Q^2)$), and transverse~($P_{T} (E_e,Q^2)$) components of the final nucleon polarization. These polarization observables provide additional and independent sensitivity to the underlying weak form factors. Most of the calculations are done assuming time-reversal~(T) invariance. 
However, we have also studied the effect of T violation on the polarization of the final nucleon for which the transverse component of the polarization i.e. $P_{T} (E_e, Q^2)$ in a direction perpendicular to the reaction plane is nonzero, provided the weak coupling constant $g_2(0)$ is purely imaginary.

In view of the difficulty in preparing the transversely polarized proton target in lab, we have performed the numerical calculations for spin asymmetries of the polarized proton assuming T invariance.
Numerical results are presented for all observables across the relevant kinematic ranges, and their sensitivities to the axial vector form factor $g_1(Q^2)$,
 as well as the weak electric form factor $g_2(Q^2)$ are systematically examined. Our results demonstrate that the combination of cross section measurements of the high-luminosity electron beams on the hydrogen target and polarization measurements may offer a relatively better framework, in comparison to using the wide energy band neutrino beam on moderate to heavy nuclear targets, for studying the axial structure of the nucleon and a feasible experiment using electron beam may provide a better insight to determine $g_{1}(Q^2)$~\cite{Ankowski:2022thw, SajjadAthar:2022pjt}.

In Sec.~\ref{cross_section}, we present the formalism to calculate the cross section for the weak charged current interaction induced electron scattering on the proton target. { Sec.~\ref{sec:spin} discusses very briefly} the formalism to calculate the spin asymmetries of the polarized target proton and the polarization components of the final nucleon assuming T invariance as well as its violation { and the detailed derivation is given in Appendix-B}. 
The results for the { total scattering cross section and the $Q^2$ distribution of the differential scattering cross section $\frac{d\sigma}{dQ^2}$ at $E_e=1.1$~GeV, which is relevant for the JLab experiment} are presented and discussed in Sec.~\ref{results:dsigma}. 
Secs.~\ref{results:pol:initial} and \ref{results:pol:final}, respectively, present the results and discussion for the longitudinal and perpendicular spin asymmetries of the initial polarized proton, and the polarization components of the final nucleon. In Sec.~\ref{results:TV}, we also discuss the transverse polarization of the final nucleon, which is nonzero in the case T invariance is violated.  { We have also performed the numerical calculations for $\frac{d\sigma}{dQ^2}$, $A_L(Q^2)$, $A_P(Q^2)$, $P_L(Q^2)$, $P_P(Q^2)$, and $P_T(Q^2)$ at $E_{e}=855$~MeV, relevant for the MAMI experiment and at $E_e=2.2$~GeV, relevant for the JLab experiment, which are qualitatively of the similar nature and therefore are presented in Supplementary Material~\cite{SM}.}
Finally, in Sec.~\ref{summary}, we conclude our findings.

\section{Cross section}\label{cross_section}
  \begin{figure}
  \begin{center}
     \includegraphics[height=3cm,width=8cm]{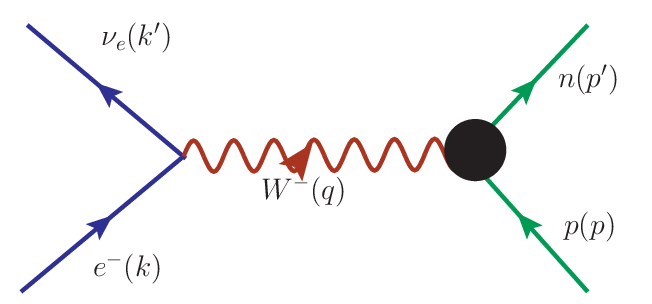}
   \caption{Feynman diagram  for the process $ e^-(k) + p(p) \rightarrow \nu_e(k^\prime) + n(p^{\prime})$.
   The quantities in the bracket represent
   four momenta of the corresponding particles.}\label{fyn_hyp}
    \end{center}
  \end{figure}
The general expression of the differential cross section for the scattering of unpolarized electron from unpolarized proton target i.e.,
\begin{eqnarray}
 e^- (k) + p (p) &\longrightarrow& \nu_e (k^\prime) + n (p^{\prime}), \label{nuc-rec} 
\end{eqnarray}
 which is diagrammatically shown in Fig.~(\ref{fyn_hyp}), in the rest frame of the initial proton, is written as:
 \begin{eqnarray}
 \label{crosv.eq}
 d\sigma&=&\frac{1}{(2\pi)^2}\frac{1}{4E_e M}\delta^4(k+p-k^\prime-p^\prime) \frac{d^3k^\prime}{2E_{k^\prime}}  
 \frac{d^3p^\prime}{2E_{p^\prime}} \overline{\sum} \sum |{\cal{M}}|^2,
 \end{eqnarray}
 where the quantities in the brackets of Eq.~(\ref{nuc-rec}) represent the four momenta of the corresponding 
 particles, $E_e$ is the electron energy, and the transition matrix element squared is expressed as:
\begin{equation}\label{matrix}
  \overline{\sum} \sum |{\cal{M}}|^2 = \frac{G_F^2 \cos^2\theta_C}{2} \cal{L}_{\alpha \beta} \cal{J}^{\alpha \beta},
\end{equation}
where $G_F$ is the Fermi coupling constant and $\theta_C$ is the Cabibbo mixing angle.
The leptonic and the hadronic tensors are given by
\begin{eqnarray}\label{L}
\cal{L}^{\alpha \beta} &=& \frac{1}{2}\mathrm{Tr}\left[\gamma^{\alpha}(1-\gamma_{5})\Lambda(k)
\gamma^{\beta}(1-\gamma_{5})\Lambda(k^\prime)~\right], \\ 
\label{J}
\cal{J}_{\alpha \beta} &=& \frac{1}{2} \mathrm{Tr}\left[\Lambda(\not{p^\prime}) J_{\alpha}
  \Lambda(\not{p}) \tilde{J}_{\beta} \right], 
\end{eqnarray} 
where the factor of $\frac{1}{2}$ in the above equations comes due to the spin averaging of the initial particles~(if both the beam and the target are unpolarized),  $\tilde{J}_{\beta} =\gamma^0 
J^{\dagger}_{\beta} \gamma^0$ with $J_{\beta}$ being the weak hadronic current. 

In Eqs.~(\ref{L}) and (\ref{J}), $\Lambda(P)=(P\!\!\!/+M_P)$ is the projection operator of an unpolarized particle with momentum $P$ and mass $M_{P}$. To take into account the spin polarization of a particle, the projection operator $\Lambda(P)$ in  Eqs.~(\ref{L}) and (\ref{J}) is replaced by the spin projection operator $\Lambda(P,s_P)$, defined as
\begin{equation}
 \Lambda (P,s_P) = (P\!\!\!/+M_P) \left(\frac{1+\gamma_5 \slashed{s}_P}{2} \right),
\end{equation}
where $s_P$ is the spin 4-vector. 
Moreover, for the polarized initial particle scattering, one has to drop the factor of $\frac{1}{2}$ in Eqs.~(\ref{L}) and (\ref{J}) and use $\Lambda (P,s_P)$ instead of $\Lambda(P)$.
Since the electron becomes highly relativistic at $E_{e}\ge150$~MeV, therefore, for the electron energies considered in the present work, the electron is fully longitudinally polarized. In this work, we have studied the total and differential scattering cross sections of the polarized electron from the unpolarized proton target and the results are presented in Sec.~\ref{results:dsigma}.

{ The hadronic current ${J}_\mu$ is expressed in terms of the vector~($V_{\mu}$) and the axial vector~($A_{\mu}$) currents as~\cite{LlewellynSmith:1971uhs}:
\begin{equation}
 {{J}}_\mu =  \bar{u} (p^\prime) (V_\mu - A_\mu) u (p)
\end{equation}
with
\begin{eqnarray}\label{vx}
 V_\mu &=& \gamma_\mu f_1(Q^2)+i\sigma_{\mu\nu} \frac{q^\nu}{2M} f_2(Q^2)
  + \frac{q_\mu}{M} f_3(Q^2),\\
  \label{vy}
  A_\mu &=&  \gamma_\mu \gamma_5 g_1(Q^2) + i \sigma_{\mu\nu}\gamma_5 \frac{q^\nu}{2M} g_2(Q^2)  
   + \frac{q_\mu} {M} g_3(Q^2) \gamma_5 , 
\end{eqnarray}
where $M$ is the mass of the nucleon. 
$q_\mu (= k_\mu - k_\mu^\prime = p_\mu^\prime -p_\mu)$ is the four momentum transfer with $Q^2 = - q^2, Q^2 >0$. $f_{1}(Q^2)$, 
$f_{2}(Q^2)$ and $f_{3}(Q^2)$ are the vector, weak magnetic and scalar form factors, and $g_{1}(Q^2)$, 
$g_{2}(Q^2)$ and $g_{3}(Q^2)$ are the axial vector, weak electric~(or induced tensor) and pseudoscalar form factors, respectively and are discussed breifly in Appendix-A. 
}
  

\section{Spin observables}\label{sec:spin}
\begin{figure}
  \begin{center}
    \hspace{-1cm}
    \includegraphics[height=6cm,width=8.5cm]{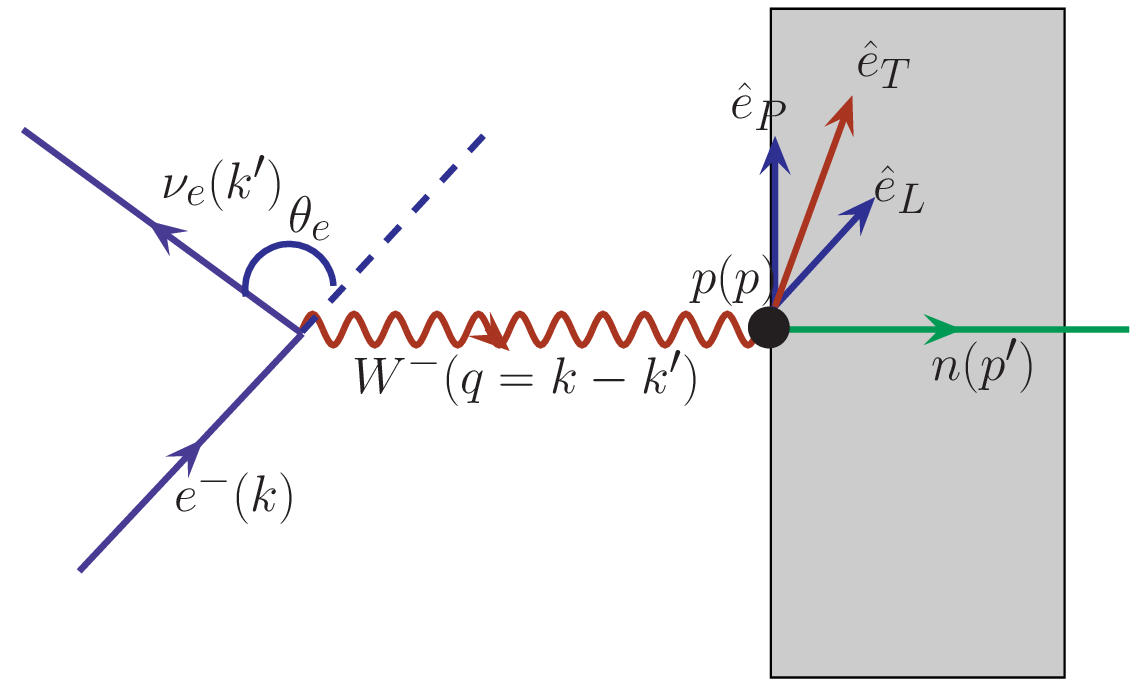}
    \includegraphics[height=6cm,width=8.5cm]{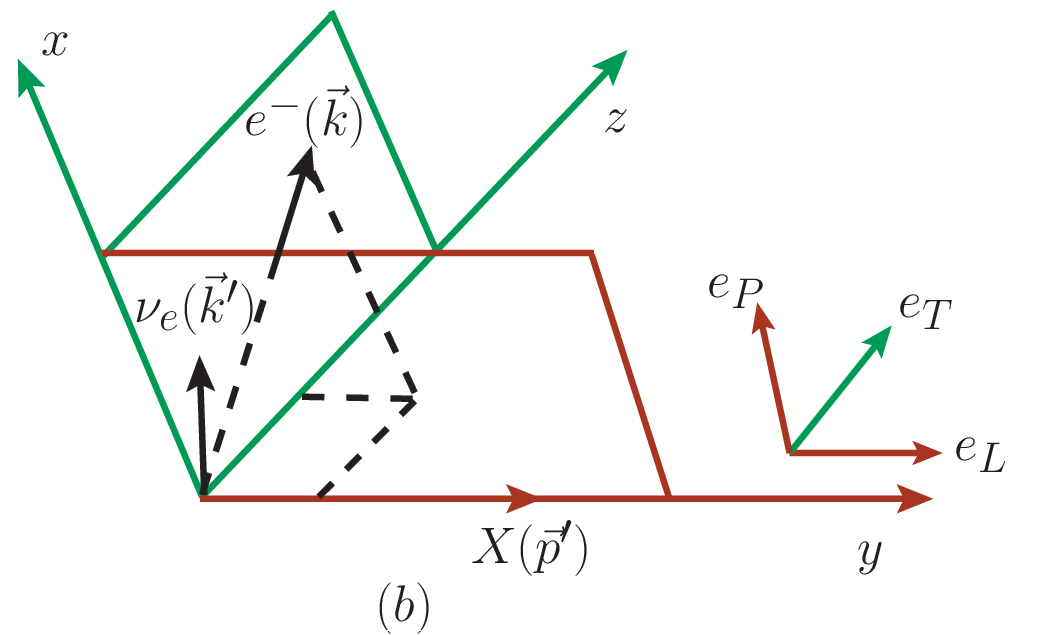}
   \caption{Diagrammatic representation of the process $ e^-(\vec{k}) + p(\vec{p}=0) \rightarrow \nu_e(\vec{k^\prime}) + 
   n(\vec{p}^{~\prime})$, and the longitudinal and perpendicular directions of the polarized proton~(left panel). The longitudinal, perpendicular and transverse directions with respect to 
  the momentum of the final nucleon~(right panel).}\label{TRI}
    \end{center}
  \end{figure}

Traditionally, the spin asymmetries are obtained using the spin projection operator formalism, for example as done by Graczyk et al.~\cite{Graczyk:2017rti, Graczyk:2023lrm} and Tomalak et al.~\cite{Tomalak:2023pdi, Borah:2024hvo}, in the case of neutrino scattering processes from the nucleon. We have used the covariant density matrix formalism following Bilenky and Christova~\cite{Bilenky:2013fra, Bilenky:2013iua, Bilekny}, which has earlier been used by us~\cite{Fatima:2018gjy, Akbar:2016awk, Akbar:2017qsf, Fatima:2018tzs, Fatima:2018wsy} for calculating the polarization components of the target nucleon. Though these are two different approaches, they lead to same results for the target asymmetries of the initial polarized nucleon. { In this work, we follow the same formalism to calculate the spin asymmetries of the polarized target and the polarization observables of the final nucleon, as developed by us earlier in Refs.~\cite{Fatima:2018gjy, Akbar:2016awk, Akbar:2017qsf, Fatima:2018tzs, Fatima:2018wsy}. The diagrammatic representation of the longitudinal, perpendicular, and transverse components of the polarized target proton as well as the polarized final nucleon if given in Fig.~\ref{TRI}. 
In the following, we have given the explicit expressions for the spin asymmetries of the polarized target and polarization components of the final nucleon
while the derivation is given in Appendix-B  for completeness.}
  
\subsection{Spin asymmetry of the polarized proton target}
\label{pol:initial}
The polarization 4-vector~($\zeta^\tau$) of the initial nucleon  is written as~\cite{Bilekny}:
\begin{eqnarray}\label{polar4:i}
\zeta^{\tau}&=&\left( g^{\tau\sigma}-\frac{p^{\tau}p^{\sigma}}{M^2}\right) \frac{  {\cal L}^{\alpha \beta}  \mathrm{Tr}
\left[\gamma_{\sigma}\gamma_{5}\Lambda(p')J_{\alpha} \Lambda(p)\tilde{J}_{\beta} \right]}
{ {\cal L}^{\alpha \beta} \mathrm{Tr}\left[\Lambda(p')J_{\alpha} \Lambda(p)\tilde{J}_{\beta} \right]}.
\end{eqnarray}

{ The expressions for the longitudinal and perpendicular components of the spin asymmetry, i.e., $A_L (Q^2)$ and $A_P (Q^2)$ are obtained, using the formalism given in Appendix-B,  as}
\begin{eqnarray}
  A_L (Q^2) &=& \frac{1}{E_e} \frac{\alpha(E_e,Q^2) |\vec{k}|^2 + \beta (E_e,Q^2) \vec{k} \cdot \vec{q}}
  {N(E_e,Q^2)},
  \label{Al} \\
 A_P (Q^2) &=& \frac{\beta(E_e,Q^2) [|\vec{q}| \sin\beta_k]}{N(E_e,Q^2) },
 \label{Ap} 
\end{eqnarray}
where $\beta_k$ is the angle between $\vec{k}$ and $\vec{q}$ and the expression for $N(E_e,Q^2) = {\cal L}_{\alpha\beta}{\cal J}^{\alpha \beta}$, $\alpha(E_e,Q^2)$, and $\beta(E_e,Q^2)$  are given in Appendix-C.

\subsection{Polarization components of the final nucleon}\label{pol:final}

{ Assuming T-invariance, the expressions for the longitudinal and perpendicular components of the final neutron polarization i.e., $P_L (Q^2)$ and $P_P (Q^2)$, obtained in the covariant density matrix formalism as discussed in Appendix-B, are expressed as}
\begin{eqnarray}
  P_L (Q^2) &=& \frac{M}{E^\prime} \frac{A(E_e,Q^2) \vec{k} \cdot \hat{p}^{\prime} + B (E_e,Q^2)
  |\vec{p}^{\,\prime}|}{N(E_e,Q^2)},
  \label{Pl} \\
 P_P (Q^2) &=& \frac{A(E_e,Q^2) [(\vec{k}.\hat{p}^{\prime})^2 - |\vec{k}|^2]}{N(E_e,Q^2) ~|\hat{p}^{\prime} \times
 \vec{k}|}.\label{Pp} 
\end{eqnarray}
The expressions for $A(E_e,Q^2)$ and $B(E_e,Q^2)$ are given in Appendix-D.

In the case of T-violation, the explicit expressions for $P_L (Q^2)$ and $P_P (Q^2)$ are the same as given in Eqs.~(\ref{Pl}) and (\ref{Pp}), except that the coefficients $A(E_e,Q^2)$ and $B(E_e,Q^2)$ are now replaced with   $A^\prime(E_e,Q^2)$ and $B^\prime(E_e,Q^2)$. 
The expression for { the transverse component of the neutron polarization} $P_T (Q^2)$, { which is nonzero in this case,} is expressed as
\begin{eqnarray} \label{Pt}
  P_T (Q^2) &=& \frac{C^\prime(E_e,Q^2) M |\vec{p}^{\,\prime}|[(\vec{k}.\hat{p}^{\prime})^2 - |\vec{k}|^2]}{N(E_e,Q^2)~
  |\hat{p}^{\prime} \times \vec{k}|}.
\end{eqnarray}
The expressions for $A^\prime(E_e,Q^2)$, $B^\prime(E_e,Q^2)$, and $C^\prime(E_e,Q^2)$ are given in Appendix-E.

\section{Results and Discussion}\label{results}
\subsection{Total and differential scattering cross sections}\label{results:dsigma}
\begin{figure}
\begin{center}
\includegraphics[width=5.5cm,height=8cm]{sigma_ga_variation.eps}
\includegraphics[width=5.5cm,height=8cm]{total_sigma_MA_variation_polarized_proton.eps}
\includegraphics[width=5.5cm,height=8cm]{total_sigma_g2_variation_polarized_proton.eps}
\caption{Total cross section $\sigma$ as a function of $E_{e}$ for the process $e^- + p \longrightarrow \nu_{e} + n$, when the initial electron is polarized. (Left panel) shows the results for the different parameterizations of $g_{1}(Q^2)$ viz. the dipole parameterization with $M_{A} = 1.026$~GeV shown by solid line, the lattice gauge parameterization of Robert and Chen~\cite{Chen:2021guo, Chen:2022odn} shown by the dash-dotted line, the $z$ expansion for the MINERvA hydrogen, LQCD, deuterium, and combined hydrogen-LQCD fits are represented by double-dot-dashed line, double dash-dotted line, dashed line and dotted line, respectively. (Middle panel) shows the results for $\sigma$ using the different values of the axial dipole mass $M_{A}$ viz. $M_{A} =1.026$~GeV~(dashed line), 1.1~GeV~(dash-dotted line), 1.2~GeV~(double-dot-dashed line), and 1.35~GeV~(double-dash-dotted line). (Right panel) shows the results for $\sigma$ obtained using the different values of $g_{2}(0)$ assuming T-invariance viz. $g_{2}^{R} (0)=0$~(solid line), +1~(dashed line), +2~(dash-dotted line), $-1$~(double-dot-dashed line), and $-2$~(double-dash-dotted line). }\label{sigma:delta1}
\end{center}
\end{figure}

\begin{figure}
 \includegraphics[width=5.9cm,height=6cm]{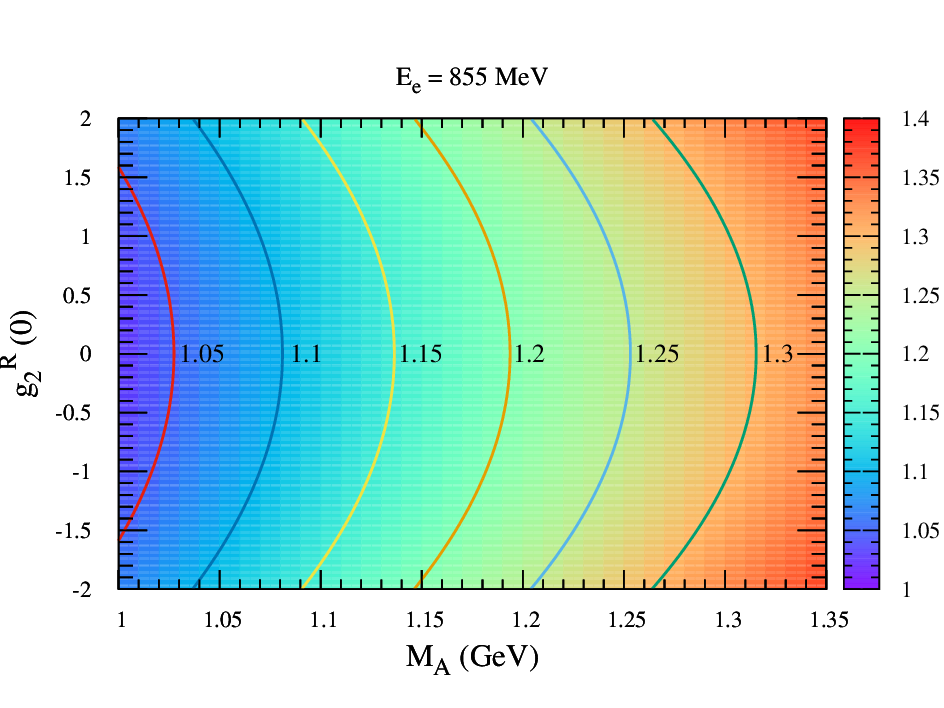}
 \includegraphics[width=5.9cm,height=6cm]{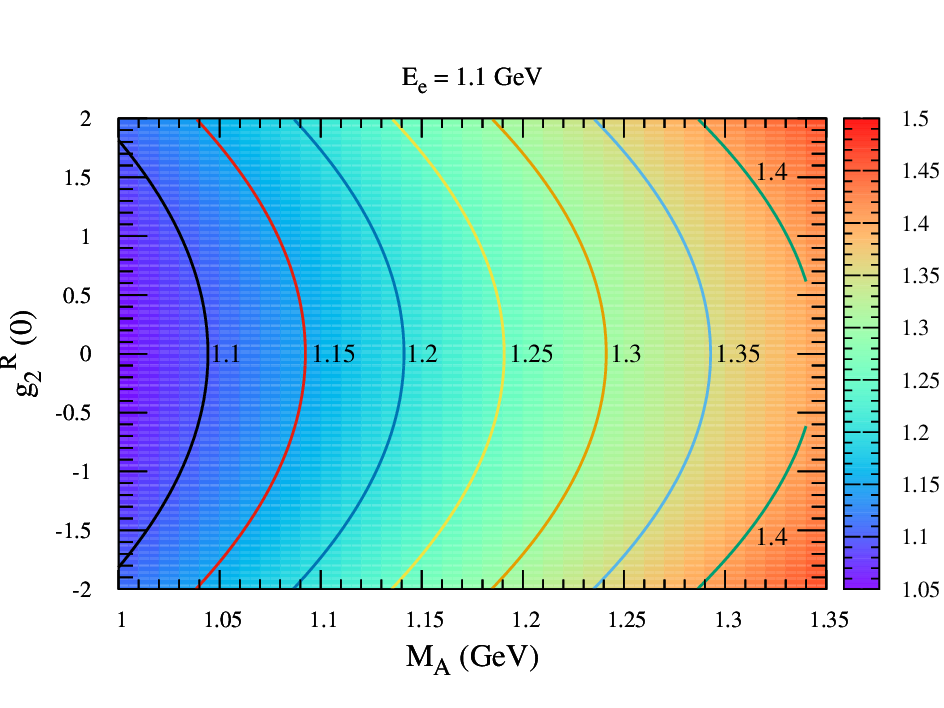}
 \includegraphics[width=5.9cm,height=6cm]{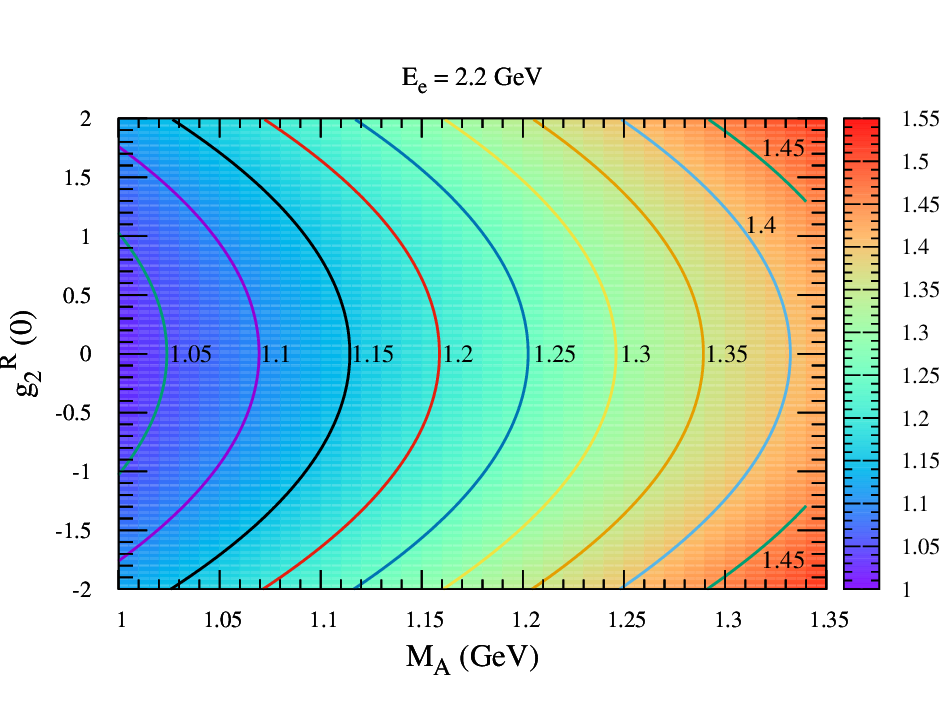}
\caption{$g_2^R(0)-M_A$ correlation for $\sigma$ for the process $e^- + p \longrightarrow \nu_{e} + n$, when the initial electron is polarized at $E_{e}=855$~MeV~(left panel), 1.1~GeV~(middle panel), and 2.2~GeV~(right panel). The solid contour lines are drawn for constant $\sigma$~($\times 10^{-38}$~cm$^2$) and their values being quoted in the plot.}\label{sigma:g2:MA:correlation}
\end{figure}

\begin{figure}
\begin{center}
\includegraphics[width=5.5cm,height=7.5cm]{dsigma_dq2_ga_variation_Ee_11GeV.eps}
\includegraphics[width=5.5cm,height=7.5cm]{dsigma_dQ2_MA_variation_proton_polarized_Ee_11GeV.eps}
\includegraphics[width=5.5cm,height=7.5cm]{dsigma_dQ2_g2_variation_proton_polarized_Ee_11GeV.eps}
\caption{$\frac{d\sigma}{dQ^2}$ as a function of $Q^2$ at $E_{e}=1.1$~GeV for the process $e^- + p \longrightarrow \nu_{e} + n$, when the initial electron is polarized for the different parameterizations of $g_{1}(Q^2)$~(left panel), different values of $M_{A}$~(middle panel), and the different values of $g_2^R(0)$~(right panel). Lines and points have the same meaning as in Fig.~\ref{sigma:delta1}.}\label{dsigma:gA}
\end{center}
\end{figure}

\begin{figure}
\begin{center}
\includegraphics[width=7.5cm,height=7.5cm]{Pl_ga_variation.eps}
\includegraphics[width=7.5cm,height=7.5cm]{Pp_ga_variation.eps}

\includegraphics[width=7.5cm,height=7.5cm]{Pl_Ee_MA_variation_polarized_proton.eps}\includegraphics[width=7.5cm,height=7.5cm]{Pp_Ee_MA_variation_polarized_proton.eps}
\caption{ $A_{L} (E_e)$~(left panel) and $A_{P} (E_e)$~(right panel) as a function of $E_{e}$ for the process $e^- + p \longrightarrow \nu_{e} + n$, when the initial proton is polarized for the different parameterizations of $g_{1}(Q^2)$~(top panel) and $M_{A}$~(bottom panel). Lines and points have the same meaning as in Fig.~\ref{sigma:delta1}.}\label{PlPp:Ee:gA:MA}
\end{center}
\end{figure}

\begin{figure}
\begin{center}
\includegraphics[width=5cm,height=7.5cm]{sigma_ga_variation_MA_band.eps}
\includegraphics[width=5cm,height=7.5cm]{Pl_ga_variation_MA_band.eps}
\includegraphics[width=5cm,height=7.5cm]{Pp_ga_variation_MA_band.eps}
\caption{$\sigma$~(left panel), $A_{L} (E_e)$~(middle panel) and $A_{P} (E_e)$~(right panel) as a function of $E_{e}$ for the process $e^- + p \longrightarrow \nu_{e} + n$, when the initial proton is polarized for the different parameterizations of $g_{1}(Q^2)$. Lines and points have the same meaning as in Fig.~\ref{sigma:delta1} and the band corresponds to the dipole parameterization with $M_{A}$ in the range 1.026--1.35~GeV.}\label{PlPp:Ee:gA:MA:band}
\end{center}
\end{figure}

\begin{figure}
\begin{center}
\includegraphics[width=7.5cm,height=7.5cm]{Pl_Ee_g2R_variation_polarized_proton.eps}
\includegraphics[width=7.5cm,height=7.5cm]{Pp_Ee_g2R_variation_polarized_proton.eps}
\caption{ $A_{L} (E_e)$~(left panel) and $A_{P} (E_e)$~(right panel) as a function of $E_{e}$ for the process $e^- + p \longrightarrow \nu_{e} + n$, when the initial proton is polarized for the different values of $g_{2}^R(0)$. Lines and points have the same meaning as in Fig.~\ref{sigma:delta1}.}\label{Pl:g2}
\end{center}
\end{figure}

\begin{figure}
\begin{center}
\includegraphics[width=5cm,height=7.5cm]{total_sigma_M2_variation_polarized_proton.eps}
\includegraphics[width=5cm,height=7.5cm]{Pl_M2_variation_polarized_proton.eps}
\includegraphics[width=5cm,height=7.5cm]{Pp_M2_variation_polarized_proton.eps}
\caption{$\sigma$~(left panel), $A_{L} (E_e)$~(middle panel) and $A_{P} (E_e)$~(right panel) as a function of $E_{e}$ for the process $e^- + p \longrightarrow \nu_{e} + n$, when the initial proton is polarized. Lines and points have the same meaning as in Fig.~\ref{sigma:delta1} and the band corresponds to the $M_{2}$ variation in the range 1.026--1.35~GeV.}\label{PlPp:Ee:M2:band}
\end{center}
\end{figure}

\begin{figure}
\begin{center}
\includegraphics[width=5.5cm,height=7.5cm]{Pl_q2_ga_variation_Ee_11GeV.eps}\includegraphics[width=5.5cm,height=7.5cm]{PL_dQ2_MA_variation_proton_polarized_Ee_11GeV.eps}
\includegraphics[width=5.5cm,height=7.5cm]{PL_dQ2_g2R_variation_proton_polarized_Ee_11GeV.eps}

\includegraphics[width=5.5cm,height=7.5cm]{Pp_q2_ga_variation_Ee_11GeV.eps}
\includegraphics[width=5.5cm,height=7.5cm]{PP_dQ2_MA_variation_proton_polarized_Ee_11GeV.eps}
\includegraphics[width=5.5cm,height=7.5cm]{PP_dQ2_g2R_variation_proton_polarized_Ee_11GeV.eps}
\caption{$A_{L} (Q^2)$~(top panel) and $A_{P} (Q^2)$~(bottom panel) as a function of $Q^2$ at $E_{e}=1.1$~GeV for the process $e^- + p \longrightarrow \nu_{e} + n$, when the initial proton is polarized for the different parameterizations of $g_{1}(Q^2)$~(left panel), different values of $M_{A}$~(middle panel), and different values of $g_2^R(0)$~(right panel). Lines and points have the same meaning as in Fig.~\ref{sigma:delta1}.}\label{PlPp:q2:ga}
\end{center}
\end{figure}

\begin{figure}
  \includegraphics[height=7.5cm,width=7.5cm]{PL_Ee_gA_variation_neutron_polarized.eps}
 \includegraphics[height=7.5cm,width=7.5cm]{PP_Ee_gA_variation_neutron_polarized.eps}
\caption{ $P_L (E_e)$~(left panel) and $P_P (E_e)$~(right panel) vs. $E_{e}$ for the process ${e^- + p \rightarrow \nu_e + n}$, when the final nucleon is polarized, using the different parameterizations of the axial vector form factor. Lines and points have the same meaning as in Fig.~\ref{sigma:delta1}.}\label{PlPp:Ee:ga:neutron}
\end{figure}

\begin{figure}
  \includegraphics[height=7.5cm,width=7.5cm]{PL_Ee_MA_variation_neutron_polarized.eps}
 \includegraphics[height=7.5cm,width=7.5cm]{PP_Ee_MA_variation_neutron_polarized.eps}
\caption{ $P_L (E_e)$~(left panel) and $P_P (E_e)$~(right panel) for the process ${e^- + p \rightarrow \nu_e + n}$, when the final nucleon is polarized, using the different values  of $M_{A}$. Lines and points have the same meaning as in Fig.~\ref{sigma:delta1}. }\label{PlPp:Ee:MA:neutron}
\end{figure}

\begin{figure}
  \includegraphics[height=7.5cm,width=7.5cm]{PL_Ee_gA_variation_neutron_polarized_MA_band.eps}
 \includegraphics[height=7.5cm,width=7.5cm]{PP_Ee_gA_variation_neutron_polarized_MA_band.eps}
\caption{ $P_L (E_e)$~(left panel) and $P_P (E_e)$~(right panel) vs. $E_{e}$ for the process ${e^- + p \rightarrow \nu_e + n}$, when the final nucleon is polarized, using the different parameterizations of the axial vector form factor. Lines and points have the same meaning as in Fig.~\ref{sigma:delta1}  and the band corresponds to the dipole parameterization with $M_A$ in the range 1.026--1.35~GeV.}\label{PlPp:Ee:ga:band:neutron}
\end{figure}

\begin{figure}
  \includegraphics[height=7.5cm,width=7.5cm]{PL_Ee_g2R_variation_neutron_polarized.eps}
 \includegraphics[height=7.5cm,width=7.5cm]{PP_Ee_g2R_variation_neutron_polarized.eps}
\caption{ $P_L (E_e)$~(left panel), $P_P (E_e)$~(middle panel), and $P_T (E_e)$~(right panel) vs. $E_{e}$ for the process ${e^- + p \rightarrow \nu_e + n}$, when the final nucleon is polarized, using the different values  of $g_{2}^{R}(0)$. Lines and points have the same meaning as in Fig.~\ref{sigma:delta1}.}\label{PlPp:Ee:g2R:neutron}
\end{figure}

\begin{figure}
  \includegraphics[height=7.5cm,width=7.5cm]{Pl_Ee_M2_variation_polarized_neutron.eps}
 \includegraphics[height=7.5cm,width=7.5cm]{Pp_Ee_M2_variation_polarized_neutron.eps}
\caption{ $P_L (E_e)$~(left panel) and $P_P (E_e)$~(right panel) vs. $E_{e}$ for the process ${e^- + p \rightarrow \nu_e + n}$, when the final nucleon is polarized, using the different values of $g_{2}^{R}(0)$. Lines and points have the same meaning as in Fig.~\ref{sigma:delta1} and the band corresponds to $M_2$ in the range 1.026--1.35~GeV.}\label{PlPp:Ee:M2:band:neutron}
\end{figure}

 \begin{figure}
\begin{center}
\includegraphics[width=5cm,height=7.5cm]{PL_Q2_gA_variation_neutron_polarized_Ee_11GeV.eps}
\includegraphics[width=5cm,height=7.5cm]{PL_Q2_MA_variation_neutron_polarized_Ee_11GeV.eps}
 \includegraphics[width=5cm,height=7.5cm]{PL_Q2_g2R_variation_neutron_polarized_Ee_11GeV.eps}

\includegraphics[width=5cm,height=7.5cm]{PP_Q2_gA_variation_neutron_polarized_Ee_11GeV.eps}
\includegraphics[width=5cm,height=7.5cm]{PP_Q2_MA_variation_neutron_polarized_Ee_11GeV.eps}
 \includegraphics[width=5cm,height=7.5cm]{PP_Q2_g2R_variation_neutron_polarized_Ee_11GeV.eps}
\caption{$P_{L}(Q^2)$~(top panel) and $P_{P}(Q^2)$~(bottom panel) as a function of $Q^2$ for the process $e^- + p \longrightarrow \nu_{e} + n$, when the final nucleon is polarized, for the different parameterizations of $g_1(Q^2)$~(left panel), different values of $M_{A}$~(middle panel), and different values of $g_2^R(0)$~(right panel) at $E_{e}=1.1$~GeV. Lines and points have the same meaning as in Fig.~\ref{sigma:delta1}.}\label{PlPp:q2:gA:neutron}
\end{center}
\end{figure}

\begin{figure}
  \includegraphics[height=7.5cm,width=5.5cm]{PL_Ee_g2_variation_neutron_polarized.eps}
 \includegraphics[height=7.5cm,width=5.5cm]{PP_Ee_g2_variation_neutron_polarized.eps}
 \includegraphics[height=7.5cm,width=5.5cm]{PT_Ee_g2_variation_neutron_polarized.eps}
\caption{ $P_L (E_e)$~(left panel), $P_P (E_e)$~(middle panel), and $P_T (E_e)$~(right panel) vs. $E_{e}$ for the process ${e^- + p \rightarrow \nu_e + n}$, when the final nucleon is polarized, using the different values  of $g_{2}^I(0)$ viz., $g_2^{I}(0) = 0$~(solid line), 1~(dashed line), and 2~(dash-dotted line).}\label{PlPpPt:Ee:g2:neutron}
\end{figure}

\begin{figure}
  \includegraphics[height=7.5cm,width=5.5cm]{PL_Ee_M2_variation_neutron_polarized.eps}
 \includegraphics[height=7.5cm,width=5.5cm]{PP_Ee_M2_variation_neutron_polarized.eps}
 \includegraphics[height=7.5cm,width=5.5cm]{PT_Ee_M2_variation_neutron_polarized.eps}
\caption{ $P_L (E_e)$~(left panel), $P_P (E_e)$~(middle panel), and $P_T (E_e)$~(right panel) vs. $E_{e}$ for the process ${e^- + p \rightarrow \nu_e + n}$, when the final nucleon is polarized, using the different values  of $g_{2}^I(0)$. Lines and points have the same meaning as in Fig.~\ref{PlPpPt:Ee:g2:neutron} and the band corresponds to $M_{2}$ in the range 1.026--1.35~GeV.}\label{PlPpPt:Ee:M2:band:neutron}
\end{figure}

 \begin{figure}
\begin{center}
\includegraphics[width=5.5cm,height=7.5cm]{PL_Q2_g2_variation_neutron_polarized_Ee_11GeV.eps}
 \includegraphics[width=5.5cm,height=7.5cm]{PP_Q2_g2_variation_neutron_polarized_Ee_11GeV.eps}
\includegraphics[width=5.5cm,height=7.5cm]{PT_Q2_g2_variation_neutron_polarized_Ee_11GeV.eps}
\caption{$P_{L}(Q^2)$~(left panel), $P_{P}(Q^2)$~(middle panel), and $P_{T} (Q^2)$~(right panel)  as a function of $Q^2$ for the process $e^- + p \longrightarrow \nu_{e} + n$, when the final nucleon is polarized, for the different values of $g_2^I(0)$ at $E_{e}=1.1$~GeV. Lines and points have the same meaning as in Fig.~\ref{PlPpPt:Ee:g2:neutron}.}\label{PlPpPt:q2:g2:neutron}
\end{center}
\end{figure}

In Fig.~\ref{sigma:delta1}, we have presented the results for the total scattering cross section~($\sigma(E_e)$) as a function of incoming electron energy~$(E_e)$ for the longitudinally polarized electron induced scattering off the unpolarized proton target. The left panel of the figure shows the results obtained using the different parameterizations of the axial vector form factor $g_{1}(Q^2)$ available in the literature and discussed in Appendix-A. The cross section is calculated using the traditional dipole parameterization for $g_{1}(Q^2)$ with $M_{A} = 1.026$~GeV and compared with the results obtained using the lattice parameterization of Chen and Roberts~\cite{Chen:2021guo, Chen:2022odn}, and the 
$z$ expansion fit done by Meyer et al.~\cite{MINERvA:2025ygc, Meyer:2026kdl} for the combined LQCD calculations of RQCD 2020~\cite{Bali:2023sdi}, NME~\cite{Park:2021ypf}, Djukanovic et al.~\cite{Djukanovic:2022wru}, ETM~\cite{Alexandrou:2023qbg}, and PNDME~\cite{Jang:2023zts}, for the experimental data of MINERvA hydrogen~\cite{MINERvA:2023avz} and earlier deuterium experiments which has been recently reanalyzed~\cite{Meyer:2016oeg} and for the combined LQCD and MINERvA hydrogen data~\cite{MINERvA:2025ygc}. 
It may be observed from the figure that the lattice parameterization of Chen and Roberts~\cite{Chen:2022odn} yields the largest value of the cross section when compared with the cross sections predicted by the other parameterizations of $g_{1} (Q^2)$ considered in this work. 
This cross section is almost 27\% larger than the dipole parameterization at $E_e=700$~MeV, and increases further with increase in electron energy to become 35\% and 43\% at $E_e = 1$~GeV, and 2~GeV, respectively. 
The results obtained with the parameterization using the $z$ expansion fit of the deuterium data are quite consistent with the results obtained with the predictions assuming the dipole parameterization and show a small decrement of about 3--5\% in the entire range of electron energies considered in this work as compared to the dipole results. In the low electron energy region, $E_{e}\sim700$~MeV, the results obtained with the $z$ expansion fit of MINERvA hydrogen data are consistent with the dipole results, however, as the electron energy increases the MINERvA fit yields a larger value for $\sigma$, which is about 5\% at $E_{e}=1$~GeV and becomes $\sim$ 10\% at $E_{e}=2$~GeV. The results obtained with the $z$ expansion fit of the LQCD calculations and combined MINERvA hydrogen-LQCD~\cite{MINERvA:2025ygc} are consistent with each other and are larger than the dipole results in the entire range of electron energy considered in this work. Quantitatively, they are larger by about 15\% at $E_{e} = 700$~MeV, which becomes 20\% and 25\% at $E_{e}=1$ and 2~GeV, respectively.

The middle panel of Fig.~\ref{sigma:delta1} shows the dependence of the total cross section  on the axial dipole mass $M_{A}$. 
The results for $\sigma (E_e)$ are obtained for the four different values of $M_A$~(1.026~GeV to 1.35~GeV). We find an appreciable dependence on the cross section $\sigma(E_e)$, when $M_A$ is varied and this dependence is, quantitatively, not the same at all the incident electron energies. With the increase in $M_A$, $\sigma(E_e)$ increases. For example at $E_e=1$~GeV, when $M_A$ is changed from 1.026~GeV to 1.1~GeV, the cross section increases by about $7\%$, in the region of $E_e=1-2$~GeV, while when $M_A$ is changed from 1.1~GeV to 1.35~GeV, the cross section increases by 20$\%$ at $E_{e}=1$~GeV, and becomes 25$\%$ at $E_e=2$~GeV.

The right panel of Fig.~\ref{sigma:delta1} shows the effect of the second class current assuming T invariance on the total cross section. 
 The results are obtained using a non-zero value of $g_2^R(0)$ in Eq.~(\ref{g2}), in the range $-2\le g_2^R(0) \le +2$, with a dipole mass $M_A=1.026$~GeV in $g_{1} (Q^2)$.  
 It must be pointed out that the cross section obtained using the positive as well as the negative values for $g_2^R(0)$ are exactly the same. Moreover, if one takes into account the imaginary values of the weak electric form factor, the results obtained for the cross sections are exactly the same as those obtained for $g_2^R(0)$. It is evident from the expression of $N(E_e,Q^2)$ given in Appendix-C, that the cross section depends only on the square of $g_2(Q^2)$, thus, the results obtained using the positive or negative values of $g_2^R(0)$, or $g_2^I(0)$ are exactly the same.
 We find that the cross section increases with the increase in $g_2^R(0)$, although this increase is only marginal as compared to the increase in cross section when a larger value of $M_A$ is used in the parameterization of $g_{1} (Q^2)$. For example, at $E_e=1$~GeV, when $g_2^R(0)$ is taken as 1 instead of zero, the increase is 1$\%$, which becomes 2$\%$ at $E_e=2$~GeV. And when $g_2^R(0)$ is taken as 2, instead of 1, there is a further increase of 4$\%$ at $E_e=1$~GeV, which becomes 8$\%$ at $E_e=2$~GeV.
 However, a nonzero $g_{2}^R (0)$ may mimic the effect of a larger value of $M_{A}$. This needs further study. Therefore, we show in Fig.~\ref{sigma:g2:MA:correlation}, a correlation plot for $g_{2}^R(0)$ vs. $M_{A}$ for the total cross section at the  three different values of electron energy viz., $E_{e}=855$~MeV, 1.1~GeV, and 2.2~GeV, relevant for the JLab and MAMI experiments. The figure shows a strong $M_{A}$ dependence on the cross section, which is about 30\% when $M_{A}$ is varied in the range [0.9, 1.35]~GeV. However, the cross section { shows little dependence} to the choice of $g_{2}^R(0)$ in the range $[-2, 2]$. Therefore, one may conclude, from this figure, that the presence of a non-zero value of $g_{2}^R(0)$, provided $|g_{2}^R(0)| \le 1$, is not going to affect the determination of $M_{A}$ or $g_{1} (Q^2)$ at JLab and MAMI experiments using the total cross section measurements.

To show the dependence of the different parameterization of $g_{1}(Q^2)$ on the differential scattering cross section $\frac{d\sigma}{dQ^2}$, we have presented { in the left panel of Fig.~\ref{dsigma:gA}}, the results of $\frac{d\sigma}{dQ^2}$ vs $Q^2$ at { $E_{e} =1.1$~GeV}, relevant for { the JLab experiment}, using the different parameterizations of $g_{1}(Q^2)$ like the dipole parameterization with $M_{A}=1.026$~GeV, $z$ expansion~\cite{MINERvA:2025ygc} for the MINERvA hydrogen data, LQCD, deuterium data, and combined hydrogen-LQCD fits, and the LQCD calculation by Chen and Roberts~\cite{Chen:2021guo, Chen:2022odn}. { We have also studied the effect of $g_{1}(Q^2)$ on $\frac{d\sigma}{dQ^2}$ at other electron energies relevant for JLab and MAMI viz. $E_e=855$~MeV and 2.2~GeV, and the results are presented in Ref.~\cite{SM}.}
It may be noticed from the { the left panel of Fig.~\ref{dsigma:gA}} that the cross section is quite sensitive to the choice of $g_{1}(Q^2)$ at all values of $E_{e}$ and $Q^2$ considered in the present work. Quantitatively, at $Q^2=1$~GeV$^{2}$ the results obtained with deuterium data fit are smaller than the the results obtained using the dipole parameterization by about 10\% at all values of electron energies. The results obtained using the $z$ expansion fit of the MINERvA data and combined MINERvA-LQCD are larger than the results obtained with the dipole parameterization by about 20\% and 45\%, respectively, at $Q^2=1$~GeV$^{2}$. The LQCD calculations by Chen and Roberts~\cite{Chen:2021guo, Chen:2022odn} yield the largest value of cross section as compared to the dipole parameterization, which is about 80\% 
at $Q^2=1$~GeV$^{2}$.

{ The middle panel of} Fig.~\ref{dsigma:gA} depicts the results for $Q^2$ distribution using different values of the axial dipole mass $M_A$ in the range 1.026--1.35~GeV, in the axial vector form factor~(Eq.~(\ref{g1})) at $E_{e}=1.1$~GeV. We find a strong dependence of $Q^2$ on the differential cross section when $M_{A}$ is varied between 1.026 and 1.35 GeV. For example,  the cross section increases by about 35\% at $Q^2=0.5$~GeV$^{2}$ when $M_A$ is varied from 1.026~GeV to 1.35~GeV, which further increases with the increase in $Q^2$ and becomes larger by 60\% at $Q^2=1$~GeV$^{2}$. 

{ The right panel of }Fig.~\ref{dsigma:gA} depicts the results for $Q^2$ distribution using the different values of $g_2^R(0)$, which is associated with the second class current form factor, in the range $[-2,2]$ in Eq.~(\ref{g2}) at $E_{e}=1.1$~GeV. Unlike the dependence of $g_1(Q^2)$ and $M_{A}$ on the cross section, we find that, the $Q^2$ distribution is not much sensitive to the variation in the value of $g_{2}^R(0)$.

\subsection{Spin asymmetries of the polarized proton target}\label{results:pol:initial}
In Figs.~\ref{PlPp:Ee:gA:MA}--\ref{PlPp:q2:ga}, we have presented the results for the longitudinal and perpendicular spin asymmetries for the polarized target proton in the reaction $e^- + \vec{p} \longrightarrow \nu_e + n$. The effect of the different parameterizations for $g_{1}(Q^2)$, $M_{A}$ variation, {$M_2$ variation,} and the effect of the nonzero $g_2 (Q^2)$ on the $Q^2$ and $E_{e}$ dependent spin asymmetries is studied. 
We show in Fig.~\ref{PlPp:Ee:gA:MA}~(top panel) the dependence of the different parameterizations of $g_{1}(Q^2)$ on the energy dependent longitudinal~($A_{L} (E_e)$) and perpendicular~($A_{P} (E_e)$) spin asymmetries of the initially polarized proton. It may be noticed from the figure that the spin asymmetries are not much  affected by the choice of the parameterization for $g_{1}(Q^2)$ except for the parameterization given by Chen and Roberts~\cite{Chen:2022odn} which predicts  a smaller value~($< 2-3\%$) for $A_{L}(E_e)$ as compared to the results obtained using the other parameterizations, especially in the region of $E_{e} \ge 1.5$~GeV. In the case of $A_{P} (Q^2)$, the dependence on the choice of the axial vector form factor is even smaller.

In the bottom panel of Fig.~\ref{PlPp:Ee:gA:MA}, the results are presented for the longitudinal~($A_L (E_e)$) and perpendicular~($A_P (E_e)$) spin asymmetries by varying $M_A$ in the range 1.026--1.35~GeV. Not much dependence has been observed specially at low $E_e$, and for $A_{P} (E_e)$, when $M_A$ is varied from 1.026~GeV to 1.35~GeV. { Moreover, it is worth mentioning that the spin asymmetries depend very weakly on the axial vector form factor $g_1(Q^2)$, unlike the cross section, which depends quite strongly on $g_1(Q^2)$.
}

{ It may be observed from Figs.~\ref{sigma:delta1} and \ref{PlPp:Ee:gA:MA} that the results obtained using a higher value of $M_A$ for both the cross section and spin asymmetries mimic the results obtained using the non-dipole $g_1(Q^2)$ parameterizations. To show this explicitly, in Fig.~\ref{PlPp:Ee:gA:MA:band}, we have presented the results for $\sigma$, $A_{L} (E_e)$, and $A_P(E_e)$ vs $E_e$ using the different non-dipole parameterizations of $g_1(Q^2)$ and the band corresponds to the dipole parameterization of $g_1(Q^2)$ with $M_A$ in the range 1.026--1.35~GeV. The results for $\sigma$, $A_{L} (E_e)$, and $A_P(E_e)$ obtained using the lattice parameterization of Chen and Roberts~\cite{Chen:2022odn} are higher than the dipole band while the results obtained with the $z$-expansion fit of the deuterium data are smaller than the dipole band. The results obtained with the $z$-expansion fit of the LQCD, MINERvA-hydrogen data and the combined hydrogen data and LQCD lie within the dipole band for all the three observables. }

Interestingly, it may be observed { from the left panel of Fig.~\ref{PlPp:Ee:gA:MA:band}} that adopting a larger axial mass, for instance $M_A=1.4$~GeV within the conventional dipole parameterization, effectively reproduces the results obtained from the parameterization of Chen et al.~\cite{Chen:2022odn} based on lattice QCD calculations. Likewise, choosing $M_A=1.3$~GeV in the dipole form closely mimics the outcomes derived from the z-expansion fit to the combined MINERvA hydrogen data and lattice QCD inputs~\cite{MINERvA:2025ygc}. This correspondence is quite revealing. It suggests that the apparent enhancement in the cross section or the associated modifications in polarization observables- can be phenomenologically interpreted, within the dipole framework, as arising from a larger effective value of $M_A$. 
In other words, the impact of more sophisticated form-factor parameterizations can, to a significant extent, be encapsulated by an upward shift in the axial dipole mass when using the traditional dipole fit.

In Fig.~\ref{Pl:g2}, we have presented the results for the longitudinal and perpendicular spin asymmetries as a function of $E_e$ by varying $g_{2}^R(0)$ in the range $[-2,2]$. We observe a significant dependence of $g_2^R(0)$ on both $A_{L} (E_e)$ and $A_{P} (E_e)$ in the entire electron energy range considered in the present work, { unlike the results for the cross sections, which depend only weakly on the variation in $g_2^R(0)$}. Moreover,  the positive and negative values of $g_2^R(0)$ yields different values for $A_{L} (E_e)$ and $A_{P} (E_e)$, which is evident from the expressions of $\alpha(E_e,Q^2)$ and $\beta(E_e,Q^2)$
given in Appendix-C. However, we observe opposite trends for $A_{L} (E_e)$ and $A_{P} (E_e)$, when positive and negative values of $g_2^R(0)$ are used in the numerical calculations, i.e., $A_{L}(E_e)$ increases when $g_{2}^{R}(0)$ is increased from 0 to +2, while $A_{P} (E_e)$ decreases. Likewise, when $g_2^R(0)$ is decreased from 0 to $-2$, the values of $A_{L}(E_e)$ decreases while the values of $A_{P} (E_e)$ increases. 
Quantitatively, when $g_{2}^R(0)$ is varied from 0 to +2,  
$A_L (E_e)$ changes by about 10$\%$ at $E_e=1$~GeV, which becomes 20$\%$ at $E_e=2$~GeV, while when $g_{2}^R(0)$ is varied from 0 to $-2$,   $A_L (E_e)$ changes by about 15$\%$ at $E_e=1$~GeV, which becomes almost 30$\%$ at $E_e=2$~GeV.
Furthermore, the effect of $g_{2}^R(0)$ is more pronounced on $A_P(E_e)$, which is about 60$\%$ at $E_e=1$~GeV, and becomes 50$\%$ at $E_e=2$~GeV, when $g_2^R(0)$ is taken as +2 instead of 0. This enhancement further increases when negative values of $g_2^R(0)$ are considered and we find an enhancement of about 65\% at $E_e = 1$~GeV, which becomes 50\% at $E_e = 2$~GeV.

{ To study the effect of the  axial dipole mass associated with the second class current form factor $M_{2}$ on $\sigma$, $A_{L} (E_e)$, and $A_P(E_e)$, in Fig.~\ref{PlPp:Ee:M2:band}, the results are presented using a fixed value of $g_2^R(0)$ viz. $g_2^R(0) = \pm1$ and $\pm2$, while varying $M_{2}$ in the range 1.026--1.35~GeV, chosen in analogy with $M_A$ variation in the axial vector form factor $g_1(Q^2)$. The calculations are performed using the world average value of $M_A$ i.e. $M_{A}=1.026$~GeV in $g_1(Q^2)$. It may be observed from the figure that, in the case of $\sigma$ and $A_P(E_e)$, the band width of the $M_2$ band increases with increase in the value of $g_2^R(0)$. For example, in the case of $\sigma$, at $g_2^R(0)=1$, we find hardly any change in the results when $M_{2}$ is varied between 1.026 and 1.35~GeV, while at $g_2^R(0)=2$, we find some dependence of $\sigma$ on $M_2$, which is $\sim 3\%$ at $E_e=1$~GeV and increase with increase in $E_e$ to become 10\% at $E_e =2$~GeV.
Moreover, in the case of $A_L(E_e)$, we find that the negative values of $g_2^R(0)$ results in
broader $M_2$ bands as compared to the positive value of $g_2^R(0)$, which is in contrast to $A_P(E_e)$, where we find broader bands for the positive values of $g_2^R(0)$.}

To study the dependence of $g_{1}(Q^2)$ on the spin asymmetries of the initial nucleon,  we have presented, in { the left panel of Fig.~\ref{PlPp:q2:ga}}, the results for $A_L(Q^2)$ and $A_{P} (Q^2)$ vs $Q^2$ at $E_{e} = 1.1$~GeV, using different parameterizations of $g_{1}(Q^2)$ available in the literature and discussed in Appendix-A. It may be noticed from the figure that the longitudinal asymmetry shows small variation in the peak region of $Q^2$ when the different parameterizations for $g_{1}(Q^2)$ are used. The largest difference in $A_{L}(Q^2)$ is observed when the calculations are performed using the parameterization of Chen and Roberts~\cite{Chen:2021guo, Chen:2022odn}, which is about 2\% in the peak region of $Q^2$ as compared to the dipole parameterization. Moreover, the variation in $A_{P} (Q^2)$ is even smaller as compared to the results obtained for $A_{L} (Q^2)$.

{ The middle panel of} Fig.~\ref{PlPp:q2:ga} shows the results for the longitudinal and perpendicular spin asymmetries of the initial nucleon at $E_{e} = 1.1$~GeV, by varying $M_{A}$ in the range 1.026--1.35 GeV. As observed in the case of the different parameterizations of $g_{1} (Q^2)$, here also the longitudinal and perpendicular spin asymmetries are not much sensitive to the choice of $M_{A}$, except in the peak region of $Q^2$.

To study the dependence of $g_{2}^{R} (0)$ variation on the longitudinal and perpendicular spin asymmetries of the initial polarized proton,  we have presented in { the right panel of} Fig.~\ref{PlPp:q2:ga}, the results for $A_L(Q^2)$ and $A_{P} (Q^2)$ vs $Q^2$ at $E_{e} = 1.1$~GeV, by taking $g_2^R(0)$ in the range $[-2,2]$. It may be observed from the figure that the spin asymmetries show strong dependence on the choice of $g_{2}^R (0)$, which is larger in the case of the perpendicular asymmetry $A_{P} (Q^2)$ than in the case of the longitudinal asymmetry $A_{L} (Q^2)$. Quantitatively, in the peak region of $Q^2$, $A_{L} (Q^2)$ increases by about 15\% at $E_e \sim$ 1~GeV, which further increases and becomes 20\% at $E_e = 2.2$~GeV, when $g_{2}^R (0)$ is increased from 0 to +2, while when $g_2^R(0)$ is decreased from 0 to $-2$, $A_{L} (Q^2)$ decreases by about 25\% at $E_e \sim$ 1~GeV, which further decreases with increase in $E_{e}$ and becomes 50\% at $E_{e}=2.2$~GeV. In the case of perpendicular asymmetry, in the peak region of $Q^2$, $A_{P}(Q^2)$ decreases by about 60\% at $E_e=1.1$~GeV, which becomes 66\% at $E_{e}=2.2$~GeV, when $g_2^R(0)$ is increased from 0 to +2, while $A_{P}(Q^2)$ increases by about 60\% at $E_e=1.1$~GeV, which becomes 50\% at $E_{e}=2.2$~GeV, when $g_2^R(0)$ is decreased from 0 to $-2$.

\subsection{Polarization observables of the final nucleon}\label{results:pol:final}
In this section, we present the results for the polarization observables of the final nucleon considering the two cases viz. T invariance and its violation. In Sec.~\ref{results:Tin}, the results are presented, assuming T invariance, for the polarization observables of the final nucleon in the reaction $e^- + p \longrightarrow \nu_e + n$, where we have studied the effect of the different parameterizations of $g_{1} (Q^2)$ as well as $M_{A}$, $g_{2}^R (0)$, { and $M_2$} variations on the polarization components. 
To study the effect of T violation on the polarization observables, in Sec~\ref{results:TV}, we have presented the results for the polarization observables of the final nucleon by varying $g_2^I (0)$ in the range 0--2 { and $M_2$ in the range 1.026--1.35~GeV}.

\subsubsection{Polarization observables in the case of T invariance}\label{results:Tin}
Fig.~\ref{PlPp:Ee:ga:neutron} shows the results obtained using the different parameterizations of $g_{1}(Q^2)$, available in the literature, for the longitudinal and perpendicular components of the neutron polarization.
In the electron energy region of $E_{e}>0.3$~GeV, we find some dependence of the choice of $g_{1}(Q^2)$ on the polarization observables, which is more pronounced in the case of $P_{P} (E_{e})$. At $E_{e} = 0.5$~GeV, $P_{L}(E_{e})$ obtained using the parameterization of Chen and Roberts~\cite{Chen:2021guo, Chen:2022odn} is almost 2\% smaller than that obtained using the dipole parameterization with $M_{A}=1.026$~GeV, and this difference further increases with the increase in electron energy, for example, it becomes 4\% at $E_{e} = 2$~GeV.
In the case of perpendicular component of the neutron polarization, at $E_{e}=0.5$~GeV, the result obtained with the parameterization of Chen and Roberts~\cite{Chen:2021guo, Chen:2022odn} is almost 15\% larger than the result obtained with the dipole parameterization with $M_{A}=1.026$~GeV, and becomes 18\% larger at $E_{e} = 2$~GeV.

To show the dependence of the polarization observables on $M_{A}$, in Fig.~\ref{PlPp:Ee:MA:neutron}, we have presented the results for $P_{L} (E_e)$ and $P_{P}(E_e)$ using the value of $M_A$ in the range 1.026--1.35~GeV. It may be noticed from the figure that the $M_{A}$ variation is more pronounced in $P_{P} (E_e)$ as compared to $P_{L} (E_e)$. Quantitatively, we find that in the energy range of $E_{e} = 1-2$~GeV, the value of $P_{L} (E_{e})$ decreases by about 2--3\% when $M_{A}$ is increased from 1.026~GeV to 1.35~GeV, while we observe an increase of about 13--15\% in the value of $P_{P}(E_{e})$ when $M_{A}$ is increased from 1.0 to 1.35GeV for $E_{e} = 1-2$~GeV.

{ In Fig.~\ref{PlPp:Ee:ga:band:neutron}, the results are presented for $P_L(E_e)$ and $P_P(E_e)$ using the different non-dipole parameterizations of $g_1(Q^2)$ and compared with the results obtained using the dipole parameterization of $g_1(Q^2)$ with $M_{A}$ being varied in the range 1.026--1.35~GeV. We find similar trends as observed in the case of $\sigma$, $A_L(E_e)$, and $A_P(E_e)$ in Fig.~\ref{PlPp:Ee:M2:band}, that the results obtained with $z$-expansion fits except the deuterium data fit lie within the dipole band while the results obtained with the lattice parameterization of Chen et al.~(deuterium data fit) are larger~(smaller) than the dipole band for both $P_L(E_e)$ and $P_P(E_e)$.}

In Fig.~\ref{PlPp:Ee:g2R:neutron}, we present the results for  $P_L( E_e)$ and $P_P(E_e)$ as a function of electron energy  $E_e$  for different values of  $g_2^R(0)$. It is evident that when $g_2^R (0)$ is varied from 0 to +2,  $P_L( E_e)$ changes by only about 10$\%$, whereas varying  $g_2^R (0)$ from 0 to $-2$ leads to a much larger change of approximately 25$\%$. In contrast, for  $P_P (E_e)$, the trend is reversed: increasing  $g_2^R (0)$ from 0 to +2 results in a significant change of about 25$\%$, while decreasing it from 0 to $-2$ produces a comparatively smaller change of around 10$\%$. This clearly highlights that  $P_L( E_e)$ and $P_P( E_e)$ exhibit opposite sensitivities to the variation of  $g_2^R (0)$.

{ To study the effect of $M_2$ variation on $P_L( E_e)$ and $P_P(E_e)$,  the results are presented in Fig.~\ref{PlPp:Ee:M2:band:neutron} for the fixed values of $g_2^R(0)$, namely $g_2^R(0)=\pm1$ and $\pm2$ while $M_2$ being varied in the range 1.026--1.35~GeV. In the case of $P_L(E_e)$, we find that the negative values of $g_2^R(0)$ yield broader $M_2$ bands than those obtained for the positive values of $g_2^R(0)$. Moreover, there is almost no effect of $M_2$ variation for $g_2^R(0)=1$ in $P_L(E_e)$. On the contrary, for $P_{P}(E_e)$, we find that the negative values of $g_2^R(0)$ lead to  broader $M_2$ bands.}

{ The left panel of} Fig.~\ref{PlPp:q2:gA:neutron} shows the results for the neutron polarization components $P_{L} (Q^2)$ and $P_{P} (Q^2)$ vs. $Q^2$ at $E_{e} = 1.1$~GeV, by using the different parameterizations of $g_{1} (Q^2)$, as discussed in Appendix-A.
 It may be observed from the figure that the neutron polarization components show almost no dependence on the choice of the axial vector
  form factor at all values of $E_{e}$ and $Q^2$ considered in the present work.

To study the effect of $M_A$ on the polarization components, in { the middle panel of Fig.~\ref{PlPp:q2:gA:neutron}}, we have presented the results for $P_{L}(Q^2)$ and $P_{P}(Q^2)$ at $E_{e} =1.1$~GeV, by varying $M_{A}$ in the range 1.026--1.35~GeV. For both
$P_{L}(Q^2)$ and $P_{P}(Q^2)$, the different values of $M_{A}$ yield similar results for the polarization components at all values of $E_{e}$ and $Q^2$.

In { the right panel of Fig.~\ref{PlPp:q2:gA:neutron}}, we present the results for  $P_L ( Q^2 )$ and  $P_P( Q^2 )$ as function of  $Q^2$ at $E_ e=1.1$~GeV, for a range of values of  $g_2^R ( 0 )$ (i.e.,   $-2\le g_2^R ( 0 ) \le$ 2). It is clearly observed that the variation is more pronounced in  $P_P( Q^2 )$  than in  $P_L ( Q^2 )$. For instance, when  $g_2^R ( 0 )$ is varied from 0 to 2,  $P_L ( Q^2 )$ changes by only about 5$\%$, whereas varying it from 0 to $-2$ results in a significantly larger change of nearly 20$\%$. This percentage change further increases with increasing  $E_ e$; moreover, the behavior of  $P_L ( Q^2 )$    itself is altered at high  $Q^2$  for  $g_2^R ( 0 )$ = 1--2. In contrast, for   $P_P( Q^2 )$, varying  $g_2^R ( 0 )$ from 0 to 2 or from 0 to $-2$ leads to changes in opposite directions, while the magnitude of the percentage variation remains comparable for equal values of  $|g_2^R ( 0 )|$. For example, the changes are about 10$\%$ and 15$\%$, respectively, and remain of similar order for   $g_2^R ( 0 ) = 1-2$.

\subsubsection{Polarization observables in the case of T violation}\label{results:TV}
In Fig.~\ref{PlPpPt:Ee:g2:neutron}, we have presented the results for $P_{L} (E_e)$, $P_{P}(E_{e})$, and $P_{T} (E_{e})$ vs $E_{e}$ using $g_{2}^I (0)$ in the range 0--2. 
The negative values of $g_2^I(0)$ yield the same results for $P_{L} (E_e)$ and $P_{P}(E_{e})$ as obtained for positive values of $g_2^I(0)$, while for $P_{T}(E_e)$,  the negative values of $g_2^I(0)$ changes sign keeping the magnitude same, therefore, the results obtained using $g_{2}^{I} (0) < 0$ are not depicted in the figure.
It should be noted from the figure that the longitudinal and perpendicular components of the final nucleon polarization are not much sensitive to the variation in $g_2^I(0)$, while the transverse component of polarization, which arises due to the T-violating effect, is quite sensitive to the variation in $g_2^I(0)$. 
The longitudinal and perpendicular components of the neutron polarization show an increment of about 2\% when $g_{2}^I(0)$ is increased from 0 to 2, at $E_{e} =0.5$~GeV, which increases with increasing electron energy and becomes 4\% and 8\%, respectively, at $E_{e} =1$ and 2~GeV. While $P_{T} (E_e)$ shows an increment of about 20\% at $E_{e} =0.5$~GeV, which becomes 30\% and 35\%, respectively at $E_{e}=1$ and 2~GeV.

{ To study the effect of $M_2$ variation on $P_{L} (E_e)$, $P_{P}(E_{e})$, and $P_{T} (E_{e})$, the results are presented in Fig.~\ref{PlPpPt:Ee:M2:band:neutron}  for fixed values of $g_2^I(0)=1$ and $2$, while $M_2$ is varied in the range 1.026--1.35~GeV. The polarization observables of the final nucleon are quite sensitive to variation in $M_2$, especially in the energy region $E_e>0.5$~GeV. Moreover, for $P_{L} (E_e)$ and $P_{P}(E_{e})$, the width of the band increases with increasing $g_2^I(0)$, whereas for $P_{T} (E_{e})$, the width of the $M_2$ band remains almost unchanged as $g_2^I(0)$ increases. }

Fig.~\ref{PlPpPt:q2:g2:neutron} shows the results for the neutron polarization components $P_{L}(Q^2)$, $P_{P}(Q^2)$, and $P_{T}(Q^2)$ as a function of $Q^2$ at $E_{e} = 1.1$~GeV,  obtained using different values of $g_{2}^I(0)$ in the range 0--2 in Eq.~(\ref{g2}). It may be observed from the figure that $P_{L}(Q^2)$ exhibits some dependence on the choice of $g_{2}^I(0)$, especially when $g_{2}^I(0) = 2$ is used in the numerical calculations. This dependence increases with increasing electron energy. However, there is almost no effect of the different choice of $g_{2}^I(0)$~($0 \le g_2^I (0) \le 2$) on $P_{P}(Q^2)$ for all values of $Q^{2}$ and $E_{e}$ considered in the present work. In contrast, $P_{T}(Q^2)$ shows a significant dependence on $g_{2}^I(0)$, particularly in the peak region of $Q^2$. The transverse polarization component $P_{T}(Q^2)$ increases by about 30$\%$ when $g_{2}^I(0)$ is increased from 0 to 2. This increase becomes more pronounced with increasing electron energy, reaching approximately 35$\%$ and 45$\%$ at $E_{e} = 1.1$~GeV and 2.2~GeV, respectively. Therefore, studying $P_{T}(Q^2)$ for a polarized neutron produced in the scattering of electron from the free proton target could be a promising alternative to investigate the second class current.

\section{Summary and conclusion}\label{summary}
In this work, we have investigated the total scattering cross section~($\sigma$), the differential cross section~($d\sigma/dQ^2$), the longitudinal~($A_L(E_e,Q^2)$) and perpendicular~($A_P(E_e,Q^2)$) spin asymmetries of the initial polarized proton as well as the longitudinal~($P_L(E_e,Q^2)$), perpendicular~($P_P(E_e,Q^2)$), and transverse~($P_T(E_e,Q^2)$) polarization components of the final neutron,
 in the weak charged current induced electron-proton scattering. This study is motivated by experiments being planned at the Thomas Jefferson National Accelerator Facility~(JLab). The calculations have been performed both under the assumption of time-reversal~(T) invariance and without imposing T invariance. Numerical results are presented for all the above mentioned observables, and their sensitivities to the axial vector and weak electric form factors are examined. 
 
 Our main observations are as follows:
\begin{itemize}
\item [(I)] Total and differential scattering cross section 
\begin{itemize}
 \item [(i)] The choice of $g_{1}(Q^2)$ used in the literature results in substantial change on the total cross section~($\sigma$). Specifically, the combined hydrogen-LQCD fit and the lattice fit by Chen et al.~\cite{Chen:2022odn} result in an increase of about 40\% and 50\%, respectively, in the cross section from the results obtained using the dipole fit with the world average value of $M_{A}$.
 
 \item [(ii)] The recommendation for the different values of $M_{A}$ obtained in the recent experiments like MINERvA, MicroBooNE, MiniBooNE, K2K, T2K, NOMAD, etc., varying in the range 1.026~GeV to 1.35~GeV, results in an increase in the total cross section by about 40\%~($M_{A}~1.026$~GeV vs. 1.35~GeV).
 
 \item [(iii)] A non-zero value of the weak electric form factor $g_2(Q^2)$, associated with the second class currents, irrespective of whether $g_2(0)$ is purely real or purely imaginary, 
 results in an increase in the total cross section, which is about 10\% for $g_{2} (0) \approx 2$ and is small around 2\% for $g_{2} (0) \approx 1$ for electron energy above 1~GeV, considered in this work.
 
 
 \item [(iv)] In the case of the $Q^2$ dependence of the differential scattering cross section $\frac{d\sigma}{dQ^2}$, we observe similar results as in the case of $\sigma$ for the different parameterization of $g_{1} (Q^2)$ as well as for the choice of the value of $M_{A}$ and $g_2 (0)$.

\end{itemize}

\item [(II)] Spin asymmetry of the initial nucleon
\begin{itemize}
 \item [(i)] The choice of the different parameterization for $g_{1} (Q^2)$ to obtain $A_{L} (E_e)$ and $A_P (E_e)$ results in not very significant change~($< 2-3\%$) in the observables.
 
 \item [(ii)] When the different values of $M_{A}$ are considered to obtain $A_{L} (E_e)$ and $A_P (E_e)$, we observe hardly any difference except for $M_{A}=1.35$~GeV vs. $M_A =1.026$~GeV, which is also $<2\%$.
 
 \item [(iii)] For a non-zero $g_{2}^R (0)$, we find strong dependence for both $A_{L} (E_e)$ and $A_P (E_e)$, specially at high electron energies considered in this work.
 
 
 \item [(iv)] The $Q^2$ dependent spin asymmetries of the initial nucleon show very little dependence on the choice of $g_{1} (Q^2)$ parameterization or on the value of $M_{A}$ considered in this work.
 
 \item [(v)] When a non-zero value of $g_{2}^R(0)$ is taken into account, both $A_{L} (Q^2)$ and $A_{P} (Q^2)$ show strong dependence on the choice of $g_{2}^R(0)$ at all values of $Q^2$ and $E_e$ considered in the present work.
\end{itemize}

\item [(III)] Polarization observables of the final nucleon
\begin{itemize}
 \item [(i)]  In the case of the perpendicular polarization $P_{P} (E_{e})$ vs. $E_{e}$, we find some dependence on the choice of $g_{1} (Q^2)$, like $P_{P} (E_{e})$ increases by $15-18\%$ when LQCD prescription of Chen et al.~\cite{Chen:2022odn} is used in comparison to the results obtained using the dipole parameterization, while this dependence is small around 2--4\%  in the case of the longitudinal polarization $P_{L} (E_{e})$. 
 
 \item [(ii)] When the different values for $M_{A}$ are used to obtain $P_{L} (E_{e})$ and $P_{P} (E_{e})$, we find that the results in the case of $P_{P} (E_{e})$ change by about 13--15\%, whereas in the case of $P_{L} (E_{e})$, the results change by about 2--3\% when $M_{A}$ is changed to $M_{A}=1.35$~GeV from $M_{A} =1.026$~GeV. 
 
 \item [(iii)] In the case of $g_2^R(0)$ variation assuming T invariance, $P_{L} (E_{e})$ and $P_{P} (E_{e})$ show a strong dependence on the second class current form factor, when a nonzero value of $g_{2}^R (0)$ is taken in the numerical calculations. 
 
 \item [(iv)] In the case of $g_2^I(0)$ variation assuming T violation, $P_{L} (E_{e})$ and $P_{P} (E_{e})$ show a mild dependence, which is about 4--8\% in the electron energy region considered in the present work, while the transverse component of polarization $P_{T} (E_{e})$ shows a considerable enhancement, which is about 30--35\% for $g_{2}^I (0) =2$ vs. 0.
 
 \item [(v)] { In the case of $M_2$ variation in $g_2^R(Q^2)$, we find both $P_{L} (E_{e})$ and $P_{P} (E_{e})$ to be sensitive to the choice of $M_{2}$. }
 
 \item [(vi)] { In the case of $M_2$ variation in $g_2^I(Q^2)$, we find that in the case of $P_{L} (E_{e})$ and $P_{P} (E_{e})$, the width of the $M_2$ band increases with increase in $g_2^I(0)$, while $P_{T} (E_{e})$ shows constant bands for the different $g_2^I(0)$ values. }
 
 \item [(vii)] When $P_{L} (Q^2)$ and $P_{P} (Q^2)$ are obtained at different electron energies, using the different parameterizations for $g_{1} (Q^2)$ or using the different values for $M_{A}$, we find hardly any change in the numerical results.
 
 \item [(viii)] We find a strong dependence of both $P_{L} (Q^2)$ and $P_{P} (Q^2)$ on $g_2^R(0)$ variation at all values of $Q^2$ and $E_{e}$ considered in the present work.
 
 \item [(ix)] When a non-zero value of $g_{2}^I(0)$ is considered, we find small change in $P_{L} (Q^2)$, which becomes almost negligible for $P_{P} (Q^2)$, whereas in the case of $P_{T} (Q^2)$, there is a substantial dependence on the choice of $g_{2}^I (0)$, which is about 30--40\% in the peak region of $Q^2$~($g_{2}^I(0) =2$ vs. 0).
\end{itemize}

\end{itemize}

Thus, to conclude the electron beam facilities at Jefferson Lab~(JLab), Mainz Microtron~(MAMI), and the upcoming Electron-Ion Collider~(EIC) offer a uniquely powerful opportunity to explore the weak nucleon form factors. With their monochromatic energy beams, precisely controlled kinematics, and exceptionally high luminosity, these facilities can deliver clean and systematically robust measurements that are difficult to achieve in neutrino experiments. Ultimately, these advances would directly empower the neutrino physics community to extract oscillation parameters with substantially improved precision and reliability, strengthening the overall scientific reach of next-generation neutrino experiments.

 \section*{Acknowledgments}
AF and MSA are thankful to the
Department of Science and Technology (DST), Government of India for providing financial assistance under Grant No.
SR/MF/PS-01/2016-AMU. 

\section*{Data Availability}
The data are not publicly available. The data are
available from the authors upon reasonable request.

\section*{Appendix-A: Weak form factors}
{ The weak vector and axial vector form factors are constrained by well-established symmetry considerations of the vector and axial vector weak currents like T-invariance, G-invariance, CVC and PCAC hypotheses, etc.~\cite{Pais:1971er, LlewellynSmith:1971uhs, Marshak, Block:1964gj, Weinberg:1958ut, Goldberger:1958vp, Bernstein}. 
In the following, we give the explicit expressions of the vector and axial vector form factors:

\begin{itemize}
 \item [(i)] The vector form factors are expressed as
\begin{equation}
 f_{i} (Q^2) = f_{i}^{p} (Q^2) - f_{i}^{n} (Q^2), \qquad \quad i=1,2
\end{equation}
where $f_{i}^{p,n} (Q^2)$ are the electromagnetic form factors of the nucleon, which, in turn, are expressed in terms of the Sachs electric 
and magnetic form factors $G_E^{p,n} (Q^2)$ and $G_M^{p,n} (Q^2)$ of the nucleons.
For $G_E^{p,n}(Q^2)$ and $G_M^{p,n}(Q^2)$ various parameterizations are available in the literature and in our 
numerical calculations, we have used the parameterization given by Bradford et al.~\cite{Bradford:2006yz}, known in the literature as BBBA05.

\item [(ii)] For the axial vector form factor $g_{1}(Q^2)$, traditionally a dipole parameterization has been used:
\begin{eqnarray}\label{g1}
 g_{1}(Q^2)=g_{A}(0)\left(1+\frac{Q^2}{M_{A}^2}\right)^{-2},
\end{eqnarray}
where $M_A$ is the axial dipole mass and $g_A(0)$ is the axial charge. For the numerical calculations, we have used the world average 
value of $M_A=1.026$ GeV~\cite{Bernard:2001rs} unless stated, and the value of $g_{A} (0)=1.267$ is determined from the neutron beta decays~\cite{ParticleDataGroup:2024cfk}.

\item [(iii)] Recently, in the literature, the axial vector form factor is determined using the $z$ expansion~\cite{Hill:2010yb}, which provides a conformal mapping that transforms $Q^2$ into a small expansion variable $z$ across the entire kinematic region relevant to quasielastic scattering. The conformal variable $z$ is defined as
\begin{equation}
 z=\frac{\sqrt{t_c + Q^2} - \sqrt{t_c - t_0}}{\sqrt{t_c + Q^2} + \sqrt{t_c - t_0}},
\end{equation}
where the parameter $t_c \le (3m_{\pi})^2$, with $m_{\pi}$ being the mass of the pion, is fixed by the particle production threshold for the axial vector current interaction, and the variable $t_0=-0.5$~GeV$^2$ fixes the value of $Q^2$ for which $z=0$ is satisfied.

Within this framework, the axial vector form factor is expressed as a power series in the variable $z$:
\begin{equation}
 g_1(z) = \sum_{k=0}^{\infty} a_k z^k.
\end{equation}
In general, the sum over $k$ is truncated at a finite order $k_{max}$, which is determined by the fit range and data accuracy. In this work, we have followed Ref.~\cite{MINERvA:2025ygc, Meyer:2026kdl} for this fitting, where the authors concluded that the best fit is obtained with $k_{max}=6$~\cite{MINERvA:2025ygc}. 

The values of the parameters $a_{k};~(k=0-6)$ for the MINERvA hydrogen, LQCD, previous deuterium, and combined MINERvA hydrogen and LQCD fits as obtained in Ref.~\cite{MINERvA:2025ygc} are tabulated in Table~\ref{tab:ak}.
\begin{table*}
\centering
\begin{tabular*}{160mm}{@{\extracolsep{\fill}} c  c c c c }\hline\hline
&MINERvA hydrogen fit& LQCD fit & deuterium fit& Combined hydrogen-LQCD fit \\ \hline
$a_0$ & 0.61490770 & 0.71742019 & 0.54264533 & 0.71070233 \\
$a_1$ & $-1.64778080$ & $-1.72089706$ & $-2.08493637$ & $-1.74307738$ \\
$a_2$ & 0.94181417 & 0.30982708 & $1.89831616$ & 0.37944565 \\
$a_3$ & 0.41239729 & 1.62125837 & 2.40319245 & 1.69894456 \\
$a_4$ & 0.36611559 & $-0.27506993$ & $-5.88979056$ & $-0.60326876$ \\
$a_5$ & $-1.18722194$ & $-1.25297945$ & 4.14554900 & $-0.95690585$ \\
$a_6$ & 0.49976799 & 0.60044079 & $-1.01497601$ & 0.51415945 \\
\hline\hline
\end{tabular*}
\caption{Values of the parameters $a_k;~(k=0-6)$ as obtained in Ref.~\cite{MINERvA:2025ygc}.}
\label{tab:ak}                                                  
\end{table*}

\item [(iv)] Chen and Roberts~\cite{Chen:2022odn} used lattice gauge formalism to determine the axial vector form factor and parameterized it as a function of $x$, where $x=Q^2/m_N^2$, with $m_N=1.18$~GeV:
\begin{equation}
 g_1(x) = \frac{1.248 + 0.039x}{1 + 1.417x + 0.318x^2 + 0.071x^3}.
\end{equation}

\begin{figure}
\begin{center}
\includegraphics[width=10cm,height=8cm]{axial_vector_FF.eps}
\caption{$g_{1}(Q^2)$ vs. $Q^2$ using the dipole parameterization with $M_{A} = 1.026~(1.35)$~GeV shown by solid lines with circle~(square). The results obtained using the lattice gauge parameterization of Chen and Roberts~\cite{Chen:2022odn} are shown by the dash-dotted line. The results obtained using the $z$ expansion~\cite{Meyer:2026kdl, MINERvA:2025ygc} for the MINERvA hydrogen, LQCD, deuterium, and combined hydrogen-LQCD fits are represented by double-dot-dashed line, double dash-dotted line, dashed line and dotted line, respectively.}\label{axial_FF}
\end{center}
\end{figure}

To study the dependence of the parameterizations of the axial vector form factor on $Q^2$, in Fig.~\ref{axial_FF}, we show the results of $g_{1}(Q^2)$ as a function of $Q^2$ using (i)~the dipole parameterization with two values of $M_{A}$ viz. $M_{A}=1.026$ and 1.35~GeV, (ii)~the $z$ expansion parameterization~\cite{MINERvA:2025ygc} using (a)~MINERvA hydrogen data, (b)~LQCD, (c)~deuterium, and (d)~combined hydrogen-LQCD fits, and (iii)~the lattice gauge parameterization given by Chen and Roberts~\cite{Chen:2022odn}. 

It may be observed from the figure that the parameterization of Chen and Roberts~\cite{Chen:2022odn} yields the largest value of $g_1(Q^2)$, while the results for the $z$ expansion of deuterium data yields the smallest value of $g_1(Q^2)$ at all values of $Q^2$. The results of the MINERvA hydrogen data for $g_{1}(Q^2)$ are consistent with the dipole parameterization with $M_{A}=1.026$~GeV in the region of $Q^2 \le 0.4$~GeV$^{2}$, however, with the increase in $Q^2$, the MINERvA data shows a larger value for $g_{1}(Q^2)$ as compared to the dipole parameterization with $M_{A} =1.026$~GeV. The results obtained with $z$ expansion of the LQCD calculations and combined MINERvA and LQCD calculations overlap in the entire range of $Q^2$ considered in this work, and these results are comparable with the results obtained with the dipole parameterization of $g_1(Q^2)$ with $M_{A}=1.35$~GeV. Thus, the results obtained with the $z$ expansion of combined LQCD and MINERvA data are almost 30\% larger than the results obtained with the traditional dipole parameterization with $M_{A}=1.026$~GeV in the entire region of $Q^2$ considered in the present work.
 
 \item [(v)] The weak electric form factor $g_2 (Q^2)$ is taken to be of dipole form, i.e., 
\begin{eqnarray}\label{g2} 
 g_{2}(Q^2)=g_{2}(0)\left(1+\frac{Q^2}{M_{2}^2}\right)^{-2},
 \end{eqnarray}
 where $M_{2}$ is the axial dipole mass corresponding to the weak electric form factor $g_2 (Q^2)$. For simplicity, in the numerical calculations, we have taken $M_{2}=M_{A}=1.026$~GeV~\cite{Fatima:2018tzs}. Moreover, throughtout this paper, whenever the variation in $M_{A}$ is studied, we have fixed $M_2$ to be 1.026~GeV and used the different values for $M_A$ in the range 1.026--1.35~GeV. Similarly, to study $M_2$ variation, we fixed $M_{A}$ to be 1.026~GeV and varied $M_{2}$ in the range 1.026--1.35~GeV.
 
There exists limited experimental information on $g_2 (0)$, which has been extracted from the analyses of nuclear $\beta$ decays~\cite{Oka:1979bw, Minamisono:2001cd, Minamisono:2011zz, Wilkinson:2000gx}, muon capture~\cite{Commins:1983ns} and the  (anti)neutrino quasielastic scattering performed at BNL~\cite{Baker:1981su, Ahrens:1988rr} and SKAT~\cite{Belikov:1983kg}. While the nuclear $\beta$ decay and muon capture experiments obatined a small value of $g_2 (0)$ namely $g_2(0) = (0.504\pm 1.134)$~\cite{Commins:1983ns} and $g_2(0) = 0 \pm 0.075$~\cite{Day:2012gb, Oka:1979bw, Minamisono:2001cd, Minamisono:2011zz, Wilkinson:2000gx},  the neutrino experiments reported limts on $g_2(0)$, which are quite large. 
For example, the neutrino induced quasielastic scattering using  bubble chamber detector~\cite{Baker:1981su} and antineutrino quasielastic scattering using scintillator detector~\cite{Ahrens:1988rr} recommended, respectively, the real values of $g_2(0)$ to be  $-3.69 \pm 1.26$ and $-1.008$. Furthermore, Belikov et al.~\cite{Belikov:1983kg} at SKAT used neutrino and antineutrino quasielastic scattering data to estimate the upper limit for $g_2(0)$ to be $-1.63$ at 90\% C.L.
Some of the earlier theoretical calculations for the (anti)neutrino-nucleon/nucleus scattering were performed assuming real as well as imaginary values of $g_2(0)$. For example; Fujii and Yamaguchi~\cite{Fujii1, Fujii2} considered both the real as well as imaginary values of $g_2(0)$ taken as 1.92, while Berman and Veltman~\cite{Berman:1964zza} used Im$g_2(0)$ to be 3.7 and 6. The best fit value for the transverse polarization calculated by Rujula and Rafael~\cite{DeRujula:1970ek} corresponds to the range $2 \le g_2(0) \le 3$. Day and McFarland~\cite{Day:2012gb} have used Re$g_2(0)$ taken as $\sim 0.2$ in their calculations. 
Therefore, following our earlier works~\cite{Fatima:2018gjy, Fatima:2018tzs, Fatima:2018wsy}, we have used $g_{2}(0)$ in the range $[-2,2]$ both for real and imaginary values, corresponding to T invariance and its violation, respectively. 
\end{itemize}
}

\section*{Appendix-B: Spin observables}
\subsection*{Appendix-B.1: Spin asymmetry of the polarized target nucleon}
{ In the rest frame of the initial nucleon, $\zeta^\tau = (0,\vec{\zeta})$ and $\vec{\zeta}$ is expressed in terms of the orthogonal vectors $\hat{e}_{i}~(i=L,P,T)$, i.e.,
 \begin{equation}\label{polarLab:i}
\vec{\zeta}=\zeta_{L} \hat{e}_{L} + \zeta_{P} \hat{e}_{P} + \zeta_{T} \hat{e}_{T},
\end{equation}
where $\hat{e}_{L}$, $\hat{e}_{P}$, and $\hat{e}_T$  are chosen to be the set of orthogonal unit vectors corresponding to the 
longitudinal, perpendicular, and transverse directions with respect to the momentum of the initial electron, shown in Fig.~\ref{TRI}, 
and are written as~\cite{Graczyk:2023lrm}:
\begin{equation}\label{vectors:i}
\hat{ e}_{L}=\frac{\vec{ k}}{|\vec{ k}|}, \qquad \quad 
\hat{ e}_{T}=\frac{\vec{ q}\times \vec{ k}}{|\vec{ q}\times \vec{ k}|}, \qquad \quad 
\hat{ e}_{P}=\frac{\vec{ k}}{|\vec{ k}|} \times \frac{\vec{ q}\times \vec{ k}}{|\vec{ q}\times \vec{ k}|} .
 \end{equation}
 Since the transverse component lies perpendicular to the reaction plane and vanishes when T invariance is assumed, therefore, the polarization vector is expressed only in terms of $\hat{ e}_{L}$ and $\hat{ e}_{P}$.
 
The longitudinal and perpendicular components of the polarization vector $\vec{\zeta}_{L,P} (Q^2)$ using Eqs.~(\ref{polarLab:i}) and (\ref{vectors:i}) may be written as
\begin{equation}\label{PL:i}
 \zeta_{L,P}(Q^2)=\vec{\zeta} \cdot \hat{e}_{L,P}~.
\end{equation}
In the rest frame of the initial nucleon, the polarization vector $\vec{\zeta}$ is expressed as
\begin{equation}\label{pol2:i}
 \vec{\zeta} = \alpha(E_e,Q^2)~ \vec{k} + \beta(E_e,Q^2)~ \vec{q},
\end{equation}
and is explicitly obtained using Eq.~(\ref{polar4:i}). The expressions for the coefficients $\alpha(E_e,Q^2)$, and $\beta(E_e,Q^2)$ are given in the Appendix~I.

The longitudinal~($A_L(Q^2)$) and perpendicular~($A_P(Q^2)$) spin asymmetries of the initial nucleon are written as:
\begin{equation}\label{PlPp:i}
 A_L (Q^2) = \zeta_L (Q^2), ~~~~~~~ A_P (Q^2) = \zeta_P (Q^2).
\end{equation}
The expressions for $A_{L}(Q^2)$ and $A_{P}(Q^2)$ are calculated using Eqs.~(\ref{PL:i}) and (\ref{pol2:i}) in Eq.~(\ref{PlPp:i}) and the explicit expressions are given in Eqs.~(\ref{Al}) and (\ref{Ap}).
}

\subsection*{Appendix-B.2: Polarization components of the final nucleon}
{   Using the covariant density matrix formalism, the polarization 4-vector~($\xi^\tau$) of the final nucleon produced in
  the reaction, given in Eq.~(\ref{nuc-rec}) is written as~\cite{Bilekny}:
\begin{eqnarray}\label{polar4}
\xi^{\tau}&=&\left( g^{\tau\sigma}-\frac{p'^{\tau}p'^{\sigma}}{{M}^2}\right) \frac{  {\cal L}^{\alpha \beta}
\mathrm{Tr}\left[\gamma_{\sigma}\gamma_{5}\Lambda(p')J_{\alpha} \Lambda(p)\tilde{J}_{\beta} \right]}
{ {\cal L}^{\alpha \beta} \mathrm{Tr}\left[\Lambda(p')J_{\alpha} \Lambda(p)\tilde{J}_{\beta} \right]}.~~~~~
\end{eqnarray}

One may write the polarization vector $\vec{\xi}$ in terms of the three orthogonal vectors $\hat{e}_{i}~(i=L,P,T)$,
{\it i.e.},
 \begin{equation}\label{polarLab}
\vec{\xi}=\xi_{L} \hat{e}_{L} + \xi_{P} \hat{e}_{P}+\xi_{T} \hat{e}_{T} ,
\end{equation}
where $\hat{e}_{L}$, $\hat{e}_{P}$ and $\hat{e}_{T}$ are chosen to be the set of orthogonal unit vectors corresponding
to the longitudinal, perpendicular and transverse directions with respect to the momentum of the final nucleon, shown in
fig~\ref{TRI}(b), and are written as
\begin{equation}\label{vectors}
\hat{ e}_{L}=\frac{\vec{ p}^{\, \prime}}{|\vec{ p}^{\, \prime}|},~~~~~
\hat{ e}_{P}=\hat{ e}_{L}\times \hat{ e}_T, ~~~~
\hat{e}_T=\frac{\vec{ p}^{\, \prime}\times \vec{ k}}{|\vec{ p}^{\, \prime}\times \vec{ k}|}.
 \end{equation}
The longitudinal, perpendicular and transverse components of the polarization vector $\vec{\xi}_{L,P,T} (Q^2)$ using
Eqs.~(\ref{polarLab}) and (\ref{vectors}) may be written as
\begin{equation}\label{PL}
 \xi_{L,P,T}(Q^2)=\vec{\xi} \cdot \hat{e}_{L,P,T}~.
\end{equation}

In the case of final nucleon polarization, we have considered two cases: (i)~when time reversal invariance is assumed, and (ii)~when time reversal is violated. As already discussed in Appendix-A, the assumption of T invariance implies all the form factors to be real, therefore, in the first case we have taken into account purely real values of $g_2(0)$ and represent it by $g_2^R(0)$. Moreover, in the second case, when time reversal is violated, the numerical calculations are performed by taking into account purely imaginary values for $g_2(0)$ and represent it by $g_2^{I} (0)$.

\subsubsection*{Appendix-B.2.1: T invariance}
In the rest frame of the initial nucleon, assuming T invariance, the polarization vector $\vec{\xi}$ is expressed as
\begin{equation}\label{pol2:TI}
 \vec{\xi} = A(E_e,Q^2)~ \vec{k} + B(E_e,Q^2)~ \vec{p}^{\, \prime} 
\end{equation}
and is explicitly calculated using Eq.~(\ref{polar4}). The expressions for the coefficients $A(E_e,Q^2)$ and $B(E_e,Q^2)$, obtained using the real values of the weak electric form factor, associated with the second class currents, i.e., $g_{2} (Q^2)=g_2^R (Q^2)$, are given in Appendix-II.

The longitudinal~($P_L(Q^2)$) and perpendicular~($P_P(Q^2)$) components of the polarization
vector in the rest frame of the final nucleon is then obtained by performing a Lorentz boost and is written
as~\cite{Fatima:2018tzs}:
\begin{eqnarray}\label{PlPp}
 P_L (Q^2) &=& \frac{M}{E^\prime} \xi_L (Q^2), \qquad \quad P_P (Q^2) = \xi_P (Q^2),
\end{eqnarray}
where $E^\prime$ is the energy of the outgoing nucleon.
The expressions for $P_L (Q^2)$ and $P_P (Q^2)$ are then obtained using Eqs.~(\ref{vectors}), (\ref{PL})
and (\ref{pol2:TI}) in Eq.~(\ref{PlPp}) and are given in Eqs.~(\ref{Pl}) and (\ref{Pp}).

\subsubsection*{Appendix-B.2.2: T violation}
To study the effect of T violation on the polarization observables of the final nucleon, 
the polarization vector $\vec{\xi}$, in the rest frame of the initial nucleon,  is expressed as
\begin{equation}\label{pol2}
 \vec{\xi} = A^\prime(E_e,Q^2)~ \vec{k} + B^\prime(E_e,Q^2)~ \vec{p}^{\, \prime} + C^\prime(E_e,Q^2)~  M (\vec{k} \times \vec{p}^{\,\prime})
\end{equation}
and is explicitly calculated using Eq.~(\ref{polar4}). The expressions for the coefficients $A^\prime(E_e,Q^2)$, $B^\prime(E_e,Q^2)$
and $C^\prime(E_e,Q^2)$ are given in Appendix-III.

The longitudinal ($P_L(Q^2)$) and perpendicular ($P_P(Q^2)$) components of the polarization
vector in the rest frame of the final nucleon are given in Eq.~(\ref{PlPp}) and the transverse~($P_T (Q^2)$) component of polarization is written
as~\cite{Fatima:2018tzs}:
\begin{eqnarray}\label{Ptt}
 P_T (Q^2) = \xi_T (Q^2).
\end{eqnarray}
In the case of T violation, the explicit expressions for $P_L (Q^2)$ and $P_P (Q^2)$ are the same as given in Eqs.~(\ref{Pl}) and (\ref{Pp}), except that the coefficients $A(E_e,Q^2)$ and $B(E_e,Q^2)$ are now replaced with   $A^\prime(E_e,Q^2)$ and $B^\prime(E_e,Q^2)$. 
The expression for $P_T (Q^2)$ is given in Eq.~(\ref{Pt}).
}

\section*{Appendix-C}\label{Appendix:N}
The expressions for $N(E_e,Q^2)$, $\alpha (E_e, Q^2)$, and $\beta(E_e,Q^2)$ are given in terms of the four form factors; $f_{1,2}(Q^2)$ and $g_{1,2}(Q^2)$ as:
\begin{eqnarray}
 N(E_e, Q^2) &=& 2f_1 (Q^2) \left[8 E_e^2 M^2-4 E_e M Q^2-2 M^2 Q^2+Q^4 \right] + f_2(Q^2) \left[4 E_e^2 Q^2-\frac{2 E_e Q^4}{M}+Q^4 \right] \nonumber \\
 &+& 2 g_1 (Q^2) \left[8 E_e^2 M^2-4 E_e M Q^2+2 M^2 Q^2+Q^4 \right] + \left|g_2 (Q^2) \right|^2 \left[4 E_e^2 Q^2-\frac{2 E_e Q^4}{M}-Q^4 \right] \nonumber \\
 &+& 4 f_1(Q^2) f_2(Q^2) Q^4 + 4f_1(Q^2) g_1(Q^2) \left[4 E_e M Q^2- Q^4 \right] + 4f_2(Q^2) g_1(Q^2) \left[4 E_e M Q^2- Q^4 \right]  \\
  \alpha (E_e,Q^2) &=& 8 f_1^2 (Q^2) M Q^2  - \frac{2 Q^4}{M} f_{2}^2 (Q^2) + 2 f_1(Q^2) f_2(Q^2) \left[4 M Q^2-\frac{Q^4}{M} \right] + 8f_1(Q^2) g_1(Q^2) M \left[4 E_e M-Q^2\right] \nonumber \\
  &+& 2 f_1(Q^2) g_2^{R}(Q^2) \left[4 E_e Q^2-\frac{Q^4}{M} \right] + 2 f_2(Q^2) g_1(Q^2) \left[\frac{Q^4}{M} - 4 E_e Q^2 \right] \nonumber \\
  &-& 2 f_2(Q^2) g_2^{R}(Q^2) \left[\frac{Q^4}{M} - 4 E_e Q^2 \right] + 2 g_1(Q^2) g_2^{R}(Q^2) \left[\frac{Q^4}{M}+4 M Q^2 \right] \\
%
  \beta(E_e,Q^2) &=& -4 f_1^2 (Q^2) M Q^2  + 4 f_{2}^2 (Q^2)E_e Q^2 + 4 g_1^2(Q^2) M (4 E_e M-Q^2) + 4f_1(Q^2) f_2(Q^2) Q^2 \left[E_e -M\right] \nonumber \\
  &+& 8 f_1(Q^2) g_1(Q^2) M (Q^2-2 E_e M) +4 f_1(Q^2) g_2^R(Q^2) \left[-4 E_e^2 M + E_e Q^2+M Q^2\right] \nonumber \\
  &+& 4 f_2(Q^2) g_1(Q^2) \left[4 E_e^2 M -E_e Q^2+M Q^2\right] - 4  f_2(Q^2) g_2^{R}(Q^2) E_e Q^2\nonumber \\
 &-& 4 g_1(Q^2) g_2^{R}(Q^2) Q^2 (E_e+M)
 \end{eqnarray} 

\section*{Appendix-D}\label{Appendix:TI}
Assuming T invariance, the expressions for the coefficients $A(E_e, Q^2)$ and $B(E_e,Q^2)$ are given by:
\begin{eqnarray}
 A(E_e, Q^2) &=& 8 f_1^2(Q^2) M Q^2 - 2 f_2^2 (Q^2)\frac{Q^4}{M} + 2 f_1 (Q^2) f_2 (Q^2) \left[4 M Q^2-\frac{Q^4}{M}\right] \nonumber \\
 &+& 8 f_1 (Q^2) g_1(Q^2) M \left[4 E_e M -Q^2\right] +  f_1 (Q^2) g_2^R (Q^2)\left[\frac{2 Q^4}{M}-8 E_e Q^2\right] \nonumber \\
 &+& 2 f_2 (Q^2) g_1 (Q^2) \left[ \frac{2 Q^4}{M}-8 E_e Q^2 \right] + 2f_2 (Q^2) g_2^{R} (Q^2) \left[ \frac{2 Q^4}{M}-8 E_e Q^2 \right] \nonumber \\
 &-& g_1 (Q^2) g_2^{R} (Q^2)\left[ \frac{2 Q^4}{M}+8 M Q^2 \right]\\
 B(E_e,Q^2) &=& 2 f_1^2(Q^2) \left[ \frac{Q^2 \left(-4 E_e M-2 M^2+Q^2\right)}{M} \right] + f_2^2 (Q^2) \left[ \frac{2 Q^4}{M}-4 E_e Q^2 \right] \nonumber \\
 &-&2 g_1^2(Q^2) \left[ \frac{\left(2 M^2+Q^2\right) \left(4 E_e M-Q^2\right)}{M} \right] - 4 f_1 (Q^2) f_2 (Q^2) \left[ \frac{Q^2 \left(3 E_e M+M^2-Q^2\right)}{M} \right] \nonumber \\
 &-& 4 f_1 (Q^2) g_1 (Q^2) \left[ \frac{\left(8 E_e^2 M^2+4 E_e \left(M^3-M Q^2\right)+Q^4\right)}{M} \right] \nonumber \\
 &+& 4 f_1 (Q^2) g_2^{R} (Q^2) \left[ -4 E_e^2 M+3 E_e Q^2+M Q^2 \right] \nonumber \\
 &+& 4 f_2(Q^2) g_1(Q^2)\left[ -4 E_e^2 M + 3 E_e Q^2-\frac{4 Q^2 \left(M^2+Q^2\right)}{M} \right] \nonumber \\
 &+& 2 f_2(Q^2) g_2^R(Q^2)\left[ \frac{ Q^2 \left(4 E_e^2 M+2 E_e \left(M^2-Q^2\right)-M Q^2\right)}{M^2} \right] + 4 Q^2 g_1(Q^2) g_2^R(Q^2) \left[E_e +M \right]
\end{eqnarray}

\section*{Appendix-E}\label{Appendix:TV}
In the case of T violation, the expressions for the coefficients $A^\prime(E_e, Q^2)$, $B^\prime(E_e,Q^2)$, and $C^\prime (E_e,Q^2)$ are given by:
\begin{eqnarray} 
 A^\prime (E_e,Q^2) &=& 8 f_1^2 (Q^2) M Q^2  -2 f_2^2 (Q^2)\frac{ Q^4}{M} +2 f_1(Q^2) f_2(Q^2) \left[4 M Q^2-\frac{Q^4}{M} \right] \nonumber \\
 &+& 8 f_1(Q^2) g_1(Q^2) M \left[4 E_e M-Q^2\right] +  f_2(Q^2) g_1(Q^2) \left[\frac{2 Q^4}{M}-8 E_e Q^2   \right]\\
B^\prime(E_e,Q^2) &=&  2 f_1^{2} (Q^2) \left[  \frac{Q^2 \left(-4 E_e M-2 M^2+Q^2\right)}{M} \right] +  f_2^2 (Q^2) \left[ \frac{2 Q^4}{M}-4 E_e Q^2 \right] \nonumber \\
&-&2 g_1^2(Q^2) \left[ \frac{ \left(2 M^2+Q^2\right) \left(4 E_e M-Q^2\right)}{M} \right] - 4 f_1(Q^2) f_2(Q^2) \left[ \frac{Q^2 \left(3 E_e M+M^2-Q^2\right)}{M} \right] \nonumber \\
&-& 4 f_1(Q^2) g_1(Q^2) \left[ \frac{ \left(8 E_e^2 M^2+4 E_e \left(M^3-M Q^2\right)+Q^4\right)}{M} \right] \nonumber \\
&+& 4 f_2(Q^2) g_1(Q^2) \left[ -4 E_e^2 M+ Q^2 (3 E_e-M)-\frac{Q^4}{M}  \right] \\
C^\prime(E_e,Q^2) &=& -\frac{4 Q^2}{M} f_1(Q^2) g_2^{I} (Q^2) -\frac{4 Q^2}{M} f_2(Q^2) g_2^{I} (Q^2) + 4 g_1(Q^2) g_2^{I} (Q^2) \left[ \frac{ \left(Q^2-4 E_e M\right)}{M} \right]
\end{eqnarray}

\newpage

\section*{Supplementary Material}
\begin{figure}[h!]
\begin{center}
\includegraphics[width=8cm,height=7.5cm]{dsigma_dq2_ga_variation_Ee_855MeV.eps}
\includegraphics[width=8cm,height=7.5cm]{dsigma_dq2_ga_variation_Ee_22GeV.eps}
\caption{$\frac{d\sigma}{dQ^2}$ as a function of $Q^2$ at $E_{e}=855$~MeV~(left panel) and 2.2~GeV~(right panel) for the process $e^- + p \longrightarrow \nu_{e} + n$, when the initial electron is polarized for the different parameterizations of $g_{1}(Q^2)$ viz. the dipole parameterization with $M_{A} = 1.026$~GeV shown by solid line, the lattice gauge parameterization of Robert and Chen~\cite{Chen:2021guo, Chen:2022odn} shown by the dash-dotted line, the $z$ expansion for the MINERvA hydrogen, LQCD, deuterium, and combined hydrogen-LQCD fits~\cite{MINERvA:2025ygc} are represented by double-dot-dashed line, double dash-dotted line, dashed line and dotted line, respectively.}\label{dsigma:gA:SM}
\end{center}
\end{figure}
\begin{figure}[h!]
\begin{center}
\includegraphics[width=8cm,height=7.5cm]{dsigma_dQ2_MA_variation_proton_polarized_Ee_855MeV.eps}
\includegraphics[width=8cm,height=7.5cm]{dsigma_dQ2_MA_variation_proton_polarized_Ee_22_GeV.eps}
\caption{$\frac{d\sigma}{dQ^2}$ as a function of $Q^2$ at $E_{e}=855$~MeV~(left panel) and 2.2~GeV~(right panel) for the process $e^- + p \longrightarrow \nu_{e} + n$, when the initial electron is polarized for the different values of $M_{A}$  viz. $M_{A} =1.026$~GeV~(dashed line), 1.1~GeV~(dash-dotted line), 1.2~GeV~(double-dot-dashed line), and 1.35~GeV~(double-dash-dotted line).}\label{dsigma:MA:SM}
\end{center}
\end{figure}
 \begin{figure}[h!]
\begin{center}
\includegraphics[width=8cm,height=7.5cm]{dsigma_dQ2_g2_variation_proton_polarized_Ee_855MeV.eps}
\includegraphics[width=8cm,height=7.5cm]{dsigma_dQ2_g2_variation_proton_polarized_Ee_22_GeV.eps}
\caption{$\frac{d\sigma}{dQ^2}$ as a function of $Q^2$ at $E_{e}=855$~MeV~(left panel) and 2.2~GeV~(right panel) for the process $e^- + p \longrightarrow \nu_{e} + n$, when the initial electron is polarized for the different values of $g_2^R(0)$ viz. $g_{2}^{R} (0)=0$~(solid line), +1~(dashed line), +2~(dash-dotted line), $-1$~(double-dot-dashed line), and $-2$~(double-dash-dotted line).}\label{dsigma:g2:SM}
\end{center}
\end{figure}
\begin{figure}[h!]
\begin{center}
 \includegraphics[width=8cm,height=7.5cm]{Pl_q2_ga_variation_Ee_855MeV.eps}
\includegraphics[width=8cm,height=7.5cm]{Pl_q2_ga_variation_Ee_22GeV.eps}

\includegraphics[width=8cm,height=7.5cm]{Pp_q2_ga_variation_Ee_855MeV.eps}
\includegraphics[width=8cm,height=7.5cm]{Pp_q2_ga_variation_Ee_22GeV.eps}
\caption{$A_{L} (Q^2)$~(top panel) and $A_{P} (Q^2)$~(bottom panel) as a function of $Q^2$ at $E_{e}=855$~MeV~(left panel) and 2.2~GeV~(right panel) for the process $e^- + p \longrightarrow \nu_{e} + n$, when the initial proton is polarized for the different parameterizations of $g_{1}(Q^2)$. Lines and points have the same meaning as in Fig.~\ref{dsigma:gA:SM}.}\label{PlPp:q2:ga:SM}
\end{center}
\end{figure}

In Figs.~\ref{dsigma:gA:SM}--\ref{dsigma:g2:SM}, the results are presented for $\frac{d\sigma}{dQ^2}$ vs. $Q^2$ at $E_e = 855$~MeV and 2.2~GeV, by using the different parameterizations of the axial vector form factor $g_1(Q^2)$, the different values of the axial dipole mass $M_A$ in the range 1.026--1.35~GeV, and the different values of $g_2^R(0)$ in the range $[-2,2]$.

In Figs.~\ref{PlPp:q2:ga:SM}--\ref{PlPp:q2:g2:SM}, the results are presented for the longitudinal and perpendicular spin asymmetries of the polarized target proton, i.e., $A_{L}(Q^2)$ and $A_{P}(Q^2)$ vs. $Q^2$ at $E_e = 855$~MeV and 2.2~GeV, by using the different parameterizations of the axial vector form factor $g_1(Q^2)$, the different values of the axial dipole mass $M_A$ in the range 1.026--1.35~GeV, and the different values of $g_2^R(0)$ in the range $[-2,2]$.

\begin{figure}
\begin{center}
\includegraphics[width=8cm,height=7.5cm]{PL_Q2_MA_variation_proton_polarized_Ee_855MeV.eps}
\includegraphics[width=8cm,height=7.5cm]{Pl_Q2_MA_variation_proton_polarized_Ee_22_GeV.eps}

\includegraphics[width=8cm,height=7.5cm]{PP_Q2_MA_variation_proton_polarized_Ee_855MeV.eps}
\includegraphics[width=8cm,height=7.5cm]{Pp_Q2_MA_variation_proton_polarized_Ee_22_GeV.eps}
\caption{$A_{L} (Q^2)$~(top panel) and $A_{P} (Q^2)$~(bottom panel) as a function of $Q^2$ at $E_{e}=855$~MeV~(left panel) and 2.2~GeV~(right panel) for the process $e^- + p \longrightarrow \nu_{e} + n$, when the initial proton is polarized for the different values of $M_{A}$. Lines and points have the same meaning as in Fig.~\ref{dsigma:MA:SM}.}\label{PlPp:q2:MA:SM}
\end{center}
\end{figure}

 \begin{figure}
\begin{center}
\includegraphics[width=8cm,height=7.5cm]{PL_Q2_g2R_variation_proton_polarized_Ee_855MeV.eps}
\includegraphics[width=8cm,height=7.5cm]{Pl_Q2_g2R_variation_proton_polarized_Ee_22_GeV.eps}

\includegraphics[width=8cm,height=7.5cm]{PP_Q2_g2R_variation_proton_polarized_Ee_855MeV.eps}
\includegraphics[width=8cm,height=7.5cm]{Pp_Q2_g2R_variation_proton_polarized_Ee_22_GeV.eps}
\caption{$A_{L} (Q^2)$~(top panel) and $A_{P} (Q^2)$~(bottom panel) as a function of $Q^2$ at $E_{e}=855$~MeV~(left panel) and 2.2~GeV~(right panel) for the process $e^- + p \longrightarrow \nu_{e} + n$, when the initial proton is polarized, using the different values of $g_2^R(0)$. Lines and points have the same meaning as in Fig.~\ref{dsigma:g2:SM}.}\label{PlPp:q2:g2:SM}
\end{center}
\end{figure}

 \begin{figure}
\begin{center}
\includegraphics[width=8cm,height=7.5cm]{PL_Q2_gA_variation_neutron_polarized_Ee_855MeV.eps}
\includegraphics[width=8cm,height=7.5cm]{PL_Q2_gA_variation_neutron_polarized_Ee_22GeV.eps}

\includegraphics[width=8cm,height=7.5cm]{PP_Q2_gA_variation_neutron_polarized_Ee_855MeV.eps}
\includegraphics[width=8cm,height=7.5cm]{PP_Q2_gA_variation_neutron_polarized_Ee_22GeV.eps}
\caption{$P_{L}(Q^2)$~(top panel) and $P_{P}(Q^2)$~(bottom panel) as a function of $Q^2$ for the process $e^- + p \longrightarrow \nu_{e} + n$, when the final nucleon is polarized, for the different parameterizations of $g_1(Q^2)$ at  $E_{e}=855$~MeV~(left panel) and 2.2~GeV~(right panel). Lines and points have the same meaning as in Fig.~\ref{dsigma:gA:SM}.}\label{PlPp:q2:gA:neutron:SM}
\end{center}
\end{figure}
In Figs.~\ref{PlPp:q2:gA:neutron:SM}--\ref{PlPp:q2:g2R:neutron:SM}, the results are presented for the longitudinal and perpendicular components of the final nucleon polarization, i.e., $P_{L}(Q^2)$ and $P_{P}(Q^2)$ vs. $Q^2$ at $E_e = 855$~MeV and 2.2~GeV, by using the different parameterizations of the axial vector form factor $g_1(Q^2)$, the different values of the axial dipole mass $M_A$ in the range 1.026--1.35~GeV, and the different values of $g_2^R(0)$ in the range $[-2,2]$.

 \begin{figure}
\begin{center}
\includegraphics[width=8cm,height=7.5cm]{PL_Q2_MA_variation_neutron_polarized_Ee_855MeV.eps}
\includegraphics[width=8cm,height=7.5cm]{PL_Q2_MA_variation_neutron_polarized_Ee_22GeV.eps}

\includegraphics[width=8cm,height=7.5cm]{PP_Q2_MA_variation_neutron_polarized_Ee_855MeV.eps}
\includegraphics[width=8cm,height=7.5cm]{PP_Q2_MA_variation_neutron_polarized_Ee_22GeV.eps}
\caption{$P_{L}(Q^2)$~(top panel) and $P_{P}(Q^2)$~(bottom panel) as a function of $Q^2$ for the process $e^- + p \longrightarrow \nu_{e} + n$, when the final nucleon is polarized, for the different values of $M_A$ at  $E_{e}=855$~MeV~(left panel) and 2.2~GeV~(right panel). Lines and points have the same meaning as in Fig.~\ref{dsigma:MA:SM}.}\label{PlPp:q2:MA:neutron:SM}
\end{center}
\end{figure}

 \begin{figure}
\begin{center}
\includegraphics[width=8cm,height=7.5cm]{PL_Q2_g2R_variation_neutron_polarized_Ee_855MeV.eps}
 \includegraphics[width=8cm,height=7.5cm]{PL_Q2_g2R_variation_neutron_polarized_Ee_22GeV.eps}

\includegraphics[width=8cm,height=7.5cm]{PP_Q2_g2R_variation_neutron_polarized_Ee_855MeV.eps}
\includegraphics[width=8cm,height=7.5cm]{PP_Q2_g2R_variation_neutron_polarized_Ee_22GeV.eps}
\caption{$P_{L}(Q^2)$~(top panel) and $P_{P}(Q^2)$~(bottom panel)  as a function of $Q^2$ for the process $e^- + p \longrightarrow \nu_{e} + n$, when the final nucleon is polarized, for the different values of $g_2^R(0)$ at  $E_{e}=855$~MeV~(left panel)  and 2.2~GeV~(right panel). Lines and points have the same meaning as in Fig.~\ref{dsigma:g2:SM}.}\label{PlPp:q2:g2R:neutron:SM}
\end{center}
\end{figure}

 \begin{figure}
\begin{center}
\includegraphics[width=8cm,height=7cm]{PL_Q2_g2_variation_neutron_polarized_Ee_855MeV.eps}
\includegraphics[width=8cm,height=7cm]{PL_Q2_g2_variation_neutron_polarized_Ee_22GeV.eps}

\includegraphics[width=8cm,height=7cm]{PP_Q2_g2_variation_neutron_polarized_Ee_855MeV.eps}
\includegraphics[width=8cm,height=7cm]{PP_Q2_g2_variation_neutron_polarized_Ee_22GeV.eps}

\includegraphics[width=8cm,height=7cm]{PT_Q2_g2_variation_neutron_polarized_Ee_855MeV.eps}
\includegraphics[width=8cm,height=7cm]{PT_Q2_g2_variation_neutron_polarized_Ee_22GeV.eps}
\caption{$P_{L}(Q^2)$~(top panel), $P_{P}(Q^2)$~(middle panel), and $P_{T} (Q^2)$~(bottom panel)  as a function of $Q^2$ for the process $e^- + p \longrightarrow \nu_{e} + n$, when the final nucleon is polarized, for the different values of $g_2^I(0)$ viz., $g_2^{I}(0) = 0$~(solid line), 1~(dashed line), and 2~(dash-dotted line), at  $E_{e}=855$~MeV~(left panel) and 2.2~GeV~(right panel).}\label{PlPpPt:q2:g2:neutron:SM}
\end{center}
\end{figure}
In Figs.~\ref{PlPpPt:q2:g2:neutron:SM}, the results are presented for the longitudinal, perpendicular, and transverse components of the final nucleon polarization, i.e., $P_{L}(Q^2)$,  $P_{P}(Q^2)$, and $P_T(Q^2)$ vs. $Q^2$ at $E_e = 855$~MeV and 2.2~GeV, by using the  different values of $g_2^I(0)$ in the range 0--2.

\end{document}